\documentclass[sigconf]{acmart}
\renewcommand\footnotetextcopyrightpermission[1]{} 

\AtBeginDocument{
  \providecommand\BibTeX{{
    \normalfont B\kern-0.5em{\scshape i\kern-0.25em b}\kern-0.8em\TeX}}}

\usepackage{amsmath,amsfonts}
\usepackage{soul}
\usepackage{algorithmic}
\usepackage{graphicx}
\usepackage{textcomp}
\usepackage{xcolor}
\usepackage{graphicx}
\usepackage{booktabs} 

\usepackage{enumitem}
\usepackage{graphicx}
\usepackage{algorithmic}
\usepackage{array}
\usepackage{mdwmath}
\usepackage{mdwtab}
\usepackage{eqparbox}
\usepackage[tight,footnotesize]{subfigure}
\usepackage{caption}
\usepackage{stfloats}
\usepackage{url}

\newcolumntype{P}[1]{>{\centering\arraybackslash}p{#1}}

\usepackage{url}
\usepackage{algorithmic}
\usepackage{algorithm2e}
\usepackage{array}
\usepackage{multirow}
\newcolumntype{P}[1]{>{\centering\arraybackslash}p{#1}}

\usepackage{verbatim}
\usepackage{tabularx}
\usepackage{subfigure}
\usepackage{caption}

\usepackage{epstopdf}
\usepackage{notoccite}
\usepackage{stackengine}

\usepackage{afterpage}

\usepackage{slashbox}

\begin{document}

\title{Black-box Model Inversion Attribute Inference Attacks on Classification Models}

\author{Shagufta Mehnaz}
\orcid{0000-0001-5850-4568}
\affiliation{%
  \institution{Dartmouth College}
  \country{United States}
}
\email{shagufta.mehnaz@dartmouth.edu}

\author{Ninghui Li}
\affiliation{%
  \institution{Purdue University}
  \country{United States}
}
\email{ninghui@purdue.edu}

\author{Elisa Bertino}
\affiliation{%
  \institution{Purdue University}
  \country{United States}
}
\email{bertino@purdue.edu}

\begin{abstract}
{Increasing use of ML technologies in privacy-sensitive domains such as medical diagnoses, lifestyle predictions, and business decisions highlights the need to better understand if these ML technologies are introducing leakages of sensitive and proprietary training data. In this paper, we focus on one kind of model inversion attacks, where the adversary knows non-sensitive attributes about instances in the training data and aims to infer the value of a sensitive attribute unknown to the adversary, using oracle access to the target classification model. We devise two novel model inversion attribute inference attacks-- confidence modeling-based attack and confidence score-based attack, and also extend our attack to the case where some of the other (non-sensitive) attributes are unknown to the adversary. Furthermore, while previous work uses accuracy as the metric to evaluate the effectiveness of attribute inference attacks, we find that accuracy is not informative when the sensitive attribute distribution is unbalanced. We identify two metrics that are better for evaluating attribute inference attacks, namely G-mean and Matthews correlation coefficient (MCC). We evaluate our attacks on two types of machine learning models, decision tree and deep neural network, trained with two real datasets. Experimental results show that our newly proposed attacks significantly outperform the state-of-the-art  attacks. Moreover, we empirically show that specific groups in the training dataset (grouped by attributes, e.g., gender, race) could be more vulnerable to model inversion attacks. We also demonstrate that our attacks' performances are not impacted significantly when some of the other (non-sensitive) attributes are also unknown to the adversary.}
\end{abstract}


\maketitle
\pagestyle{plain}
\section{Introduction}
\label{sec:intro}
Across numerous sectors, a variety of institutions are streamlining their processes by adopting machine learning (ML) technologies and leveraging commercial ML-as-a-service APIs. In many cases, these ML technologies are trained on proprietary and sensitive datasets, e.g., in the domains of personalized medicine \cite{medicine1, medicine2, medicine3, medicine4}, product recommendation~\cite{recom1, recom2, recom3}, finance and law~\cite{finlaw1, finlaw2, finlaw3}, social media~\cite{social1, social2, social3}, etc. Moreover, with the increasing use of ML technologies in personal data, we have seen a recent surge of serious privacy concerns that were previously ignored. Therefore, it is very important to also understand whether public access to such trained models introduces new attack vectors against the privacy of these proprietary and sensitive datasets used for training ML models, such as lifestyle surveys, genetic data, purchase history of sensitive items, etc.

A \emph{model inversion attack} is the one that turns the one-way journey from training data to model into a two-way one, i.e., this attack allows an adversary to infer part of the training data even when given only oracle access to the target classification model. This attack can take the most catastrophic form when the datasets used to train such machine learning models are privacy sensitive or proprietary. Recently, Fredrikson et al.~\cite{FredriksonCCS2015, PharmaUSENIX2014} proposed two formulations of model inversion attacks. In the first one, which we call \textbf{model inversion attribute inference (MIAI)} attack, the adversary aims to learn some sensitive attribute of an individual whose data are used to train the target model, and whose other attributes are known to the adversary.  This can be applied, e.g., when each instance gives information of one individual.  In the second formulation, which we call \textbf{typical instance reconstruction (TIR)} attack, the adversary is given oracle access to a classification model and a class, and aims to come up with a typical instance for that class.  For example, the adversary, when given access to a model that recognizes different users' faces, tries to reconstruct an image that is similar to a target individual's actual facial image. 

Several recent works study TIR attacks~\cite{avodji2019gamin,YangCCS2019,Zhang_2020_CVPR}.  Note that for TIR attacks to be considered successful, it is not necessary for a reconstructed instance to be quantitatively close to any specific training instance.  Evaluation is typically done by having humans assessing similarity of the reconstructed instances (e.g., reconstructed facial images) to training instances.  Thus a model that is able to learn the essence of each class and generalizes well (as opposed to relying on remembering information specific to training instances) will likely remain vulnerable to such an attack.  Indeed, it is proven~\cite{Zhang_2020_CVPR} that a model's predictive power and its vulnerability to such TIR attacks are two sides of the same coin. This is because highly predictive models are able to establish a strong correlation between features and labels and this is the property that an adversary exploits to mount the TIR attacks~\cite{Zhang_2020_CVPR}.  In other words, the existence of TIR attacks is a feature of good classification models, although the feature may be  undesirable in some settings.  We point out that such is not the case for MIAI attacks, which is evaluated by the ability to predict exact attribute values of individual instances.

In this paper, we focus only on MIAI attacks on classification models where data about individuals are used.  More specifically, we consider the attribute inference attacks where the adversary leverages black-box access to an ML model to infer the sensitive attributes of a target individual.  While attribute inference in other contexts have been studied extensively in the privacy literature, there exists little work studying to what extent model inversion introduces new AI vulnerabilities. In the rest of the paper, we refer to MIAI attacks whenever we use the term model inversion attack.

\textbf{Proposed new model inversion attacks:} 
We design two new black-box model inversion attacks: 
(1) confidence modeling-based attack, and 
(2) confidence score-based attack. 
These attacks are different in terms of the adversary's capability assumptions, and therefore, represent a variety of adversaries that can pose different levels of threats to training data privacy. We define a threat model to clearly state the adversary assumptions not only for our proposed attacks but also for the existing attacks in the literature. 
Our confidence score-based attack assumes an adversary who can only access the predicted label and the confidence scores returned by the model, whereas the adversary assumed in our confidence modeling-based attack has access to a dataset collected from the same population the target model training dataset has been obtained from. However, there is no overlap
between these two datasets. While all the existing attacks~\cite{FredriksonCCS2015, PharmaUSENIX2014} assume that the adversary has full knowledge of other non-sensitive attributes of the target individual, it is not clear how the adversary would perform in a setting where it has only partial knowledge of those attributes. To understand the vulnerability of model inversion attacks in such practical scenarios, we also propose an attack that works even when some non-sensitive attributes are unknown to the adversary. Moreover, we also investigate if there are scenarios when  model inversion attacks do not threaten the privacy of the overall dataset but are effective on some specific groups of instances (e.g., individuals grouped by race, gender, education level, etc.).
We empirically show that there exists such discrimination across different groups of the training dataset where a group is more vulnerable than the others. 

While the existing MIAI attacks~\cite{FredriksonCCS2015} have been evaluated only on decision tree models, we evaluate our attacks also on \emph{deep neural network} models. Therefore, we train two models-- a \emph{decision tree} and a \emph{deep neural network} with each of the two real datasets in our experiments, General Social Survey (GSS)~\cite{gss} and Adult dataset~\cite{adult}, to evaluate our proposed attacks.

\textbf{Effective evaluation of model inversion attacks:} To understand if model inversion attacks pose a broader risk, it is important that we use appropriate metrics to evaluate our proposed attacks as well as the existing attacks. Although the  Fredrikson  et al.  attack~\cite{FredriksonCCS2015}  primarily  uses  accuracy, in this paper, we argue that accuracy is not the best measure. This is because simply predicting the majority class for all the instances can achieve very high accuracy which certainly misrepresents the performances of model inversion attacks. Moreover, we argue that the F1 score, a widely used metric, is also not sufficient by itself since it emphasizes only the positive class, and simply predicting the positive class for all the instances can achieve significant F1 score. Hence, we propose to use G-mean~\cite{Sun2009ClassificationOI} and Matthews correlation coefficient (MCC)~\cite{MATTHEWS1975442} as metrics to design a framework that can effectively evaluate any model inversion attack. 

\textbf{Understanding model inversion attacks in-depth}: We also propose to compare the performances of various model inversion attacks with those from attacks that do not query the target model, e.g., randomly guessing the sensitive attribute according to some distribution.  Such random guessing attacks do not even access the model and thus certainly cannot invert the model. When a particular model inversion attack deployed against a target model performs similarly to such attacks, we can conclude that the target model in not vulnerable to that particular model inversion attack. Hence, in this paper, we address the following general research question- is it possible to identify when a model should be classified as vulnerable to such model inversion attacks? \emph{More specifically, does black-box access to a particular model really help the adversary to estimate the sensitive attributes in the training dataset which is otherwise impossible for the adversary to estimate (i.e., without access to the black-box model)? }

\textbf{Summary of contributions:} In summary, this paper makes the following contributions:
\begin{enumerate}
    \item We define the various capabilities of the adversary and provide a detailed threat model.  Based on the threat model we design two new black-box model inversion attacks: (1) confidence modeling-based attack and (2) confidence score-based attack. We propose to use the G-mean and Matthews correlation coefficient (MCC) metrics along with accuracy and F1 score to compare our proposed attacks with existing~\cite{FredriksonCCS2015} and other baseline attacks. 
    \item  We conduct extensive evaluation of our attacks using two types of ML models, decision tree and deep neural network, trained with two real datasets.  Evaluation results show that our proposed attacks significantly outperform the existing attacks.  
    \item We also empirically show that a particular subset of the training dataset (grouped by attributes, such as, gender, education level, etc.) could be more vulnerable than others to the model inversion attacks. 
    \item We evaluate the risks incurred by model inversion attacks when the adversary does not have knowledge of all other non-sensitive attributes of the target individual and demonstrate that our attack's performance is not impacted significantly in those circumstances.
\end{enumerate}
\begin{table*}[h]
  \centering
  \caption{Assumption of adversary capabilities/knowledge for different attack strategies.}
  \resizebox{1\textwidth}{!}{
  \begin{tabular}{ | l | c | c | c | c | c | c | c | c |}
    \hline
    \multirow{3}{*}{Attack strategy} & Predicted   & Confidence score & Target individuals' & All possible & Marginal prior of & Marginal prior  & Confusion      & Attacker \\ 
                    & label  & along with & all other (non-  & values of the & the sensitive    & of all other (non-  & matrix of  & dataset \\ 
                    &  & predicted label & sensitive) attributes & sensitive attribute & attribute & sensitive) attributes & the model  & $DS_A$\\\hline 
    Naive attack     &     &   &      &\checkmark &\checkmark &   &      &\\ \hline
    Random guessing attack   &    &    &     &\checkmark &\checkmark(optional) &   &   &      \\ \hline
    Fredrikson et al. Attack~\cite{FredriksonCCS2015} & \checkmark & & \checkmark &\checkmark &  \checkmark & \checkmark  &  \checkmark     &\\ \hline
    Confidence modeling-based attack &  \checkmark & \checkmark   &\checkmark &\checkmark &  &   &  & \checkmark \\ \hline
    Confidence score-based attack & \checkmark & \checkmark   &\checkmark &\checkmark &  &   &  &  \\ \hline
  \end{tabular}
  }
\label{table:adversary_assumptions}
\end{table*}

\section{Problem Definition and Existing Attack Strategies}
\subsection{Model Inversion Attribute Inference Attack}
\label{subsec:mia}
An ML model can be represented using a deterministic function $f: \mathcal{R}^d \longrightarrow \mathcal{Y}$. The input of this function is a $d$-dimensional vector $\mathbf{x} = [x_1, x_2, ..., x_d] \in \mathcal{R}^d$ that represents $d$ attributes and $y' \in \mathcal{Y}$ is the output. In the case of a regression problem, $\mathcal{Y} = \mathcal{R}$. However, in this work, we focus on classification problems. Therefore, more specifically, $f: \mathcal{R}^d \longrightarrow \mathcal{R}^m$ where $m$ is the number of unique class labels ($y_1, y_2, ..., y_m$) and $\mathcal{R}^m$ represents the confidence scores returned by the classification model for these $m$ class labels. Finally, the class label with the highest confidence score is considered as the output of the prediction model. We denote the dataset on which the $f$ model is trained as $DS_T$.
From now on, we use the term $y$ to represent the actual value in the training dataset $DS_T$ whereas $y'$ is used to represent the model output $f(\mathbf{x})$. The values of $y$ and $y'$ are the same in the case of a correct prediction or different in the case of an incorrect prediction by  $f$.

Now, some of the attributes in $\mathbf{x}$ introduced above could be privacy sensitive. Without loss of generality, let's assume that $x_1 \in \mathbf{x}$ is a sensitive attribute
that the individual corresponding to a data record in the training dataset does not want to reveal to the public. However, a model inversion attack may allow an adversary to infer this $x_1$ attribute value of a target individual given some specific capabilities, such as, access to the black-box model (i.e., target model), background knowledge about the target individual, etc.

\subsection{Threat Model}
\label{subsec:threat_model}
The adversary is assumed to have all or a subset of the following capabilities/knowledge (see Table~\ref{table:adversary_assumptions}):
\begin{itemize}
    \item Access to the black-box target model, i.e., the adversary can query the model with $\mathbf{x} = [x_1, x_2, ..., x_d]$ and obtain a class label $y'$ as the output. 
    \item The confidence scores returned by the target model for $m$ class labels, i.e.,  $\mathcal{R}^m$. 
    \item Full/partial knowledge of the non-sensitive attributes of the target individual except his/her sensitive attribute.
    \item All possible ($k$) values of the sensitive attribute $x_1$.
    \item Knowledge of marginal prior of the sensitive attribute $x_1$, i.e., $\mathbf{p_1} = \{p_{1,1}, p_{1,2}, ..., p_{1,k}\}$ where $k$ is the number of all possible values of $x_1$ and $p_{1,k}$ is the probability of the $k-$th unique possible value. 
    \item Knowledge of confusion matrix ($\mathcal{C}$) of the model where $\mathcal{C}[y, y']$ = $Pr[f(x) = y' | y$ is the true label$]$.   
    \item Access to a dataset collected from the same population the target model's training dataset has been obtained from. However, there is no overlap between the target model's training dataset ($DS_T$) and the dataset that the adversary has access to ($DS_A$). 
\end{itemize}

Note that, for the attacks designed in this paper, the adversary does not need the knowledge of marginal priors of any attributes (sensitive or non-sensitive) or the confusion matrix. Also, we only consider a passive adversary that does not aim to corrupt the machine learning model or influence its output in any way.

\subsection{Baseline Attack Strategies}
\label{subsec:existing_attacks}
\subsubsection{Naive Attack}
A naive model inversion attack assumes that the adversary has knowledge about the probability distribution (i.e., marginal prior) of the sensitive attribute and always predicts the sensitive attribute to be the value with the highest marginal prior. Therefore, this attack does not require access to the target model. 
Note that this attack can still achieve significant accuracy if the sensitive attribute is highly unbalanced, e.g., if the sensitive attribute can take only two values and there is an 80\%-20\% probability distribution, predicting the value with higher probability would result in 80\% accuracy.

\subsubsection{Random Guessing Attack}
The adversary in this attack also does not require access to the target model. The adversary randomly predicts the sensitive attribute by setting a probability for each possible value. The adversary may or may not have access to the marginal priors of the sensitive attribute. Fig.~\ref{fig:random_30} in Appendix~\ref{appn:random} shows the optimal performance of random guessing attack in terms of different metrics when the adversary sets different probabilities for predicting the positive class sensitive attribute independent of its knowledge of marginal prior $0.3$. Note that, predicting the positive class for all the instances with this attack (i.e., setting a probability 1 for the positive class) would result in a significantly high F1 score, mainly due to a recall of $100\%$ (Fig.~\ref{fig:random_30} in Appendix).

\subsection{Fredrikson et al. Attack~\cite{FredriksonCCS2015}}
\label{subsec:ccs15}
The Fredrikson et al.~\cite{FredriksonCCS2015} black-box model inversion attack assumes that the adversary can obtain the model's predicted label, has knowledge of all the attributes of a targeted individual (including the true $y$ value) except the sensitive attribute, has access to the marginal priors of all the attributes, and also to the confusion matrix of the target model. The adversary can query the target model multiple times by varying the sensitive attribute ($x_1$) and obtain the predicted $y'$ values.
After querying the model $k$  times with $k$ different $x_1$ values ($x_{1,0}, x_{1,1}, \ldots, x_{1,k-1}$) while keeping the other known attributes unchanged, the adversary computes $\mathcal{C}[y, y'] * p_{1,i}$ for each possible sensitive attribute value, where
\begin{equation}
\footnotesize
{
\mathcal{C}[y, y'] = Pr[f(x) = y' |\;y\;is\;the\;true\;label] \nonumber
}
\end{equation}
and $p_{1,i}$ is the marginal prior of the i-th possible sensitive attribute value. Finally, the attack predicts the sensitive attribute value for which the computed $\mathcal{C}[y, y'] * p_{1,i}$ value is the maximum.

\section{Metrics for Evaluating the Vulnerability of a Model to Inversion Attacks}
\label{sec:indepth}
Though the impact of model inversion attacks can be overwhelming, in this section, we aim to take a deep dive to understand if it is possible to determine when a model should be classified as vulnerable and if the metrics considered in the existing model inversion attack research are sufficient. More specifically, we investigate the following general research question- \emph{does black-box access to a particular model really help the adversary to estimate the sensitive attributes in the training dataset which is otherwise impossible for the adversary to estimate (i.e., without access to that black-box model)?} 

Understanding a model's vulnerability to inversion attacks requires a meaningful metric to evaluate and compare different model inversion attacks. The Fredrikson et al. attack~\cite{FredriksonCCS2015} primarily uses accuracy. However, if we care only about accuracy, the naive attack of simply guessing the majority class for all the instances can achieve very high accuracy. Another widely used metric is the F1 score. However, the F1 score of the positive class emphasizes only on that specific class and thus, as a one-sided evaluation, cannot be considered as the only metric to evaluate the attacks. Otherwise, always guessing the positive class may achieve similar or even better F1 score (mainly due to a recall of $100\%$) than any sophisticated model inversions attack that identifies the positive class instances strategically.

To understand whether access to the black-box model considerably contributes to attack performance and also to compare the baseline attack strategies (that do not require access to the model, i.e., naive attack and random guessing attack) to our proposed attacks, we use the following two metrics in addition to accuracy and F1 scores: G-mean~\cite{Sun2009ClassificationOI} and Matthews correlation coefficient (MCC)~\cite{MATTHEWS1975442}, as described below.

\textbf{G-mean}: G-mean is the geometric mean of sensitivity and specificity~\cite{Sun2009ClassificationOI}. Thus it takes all of true positives (TP), true negatives (TN), false positives (FP), and false negatives (FN) into account. With this metric, the random guessing attack can achieve a maximum performance of $50\%$. Note that, even if the adversary has knowledge of marginal priors of the sensitive attribute, it is not able to achieve a G-mean value of more than $50\%$ by setting different probabilities for predicting the positive class sensitive attribute (Fig.~\ref{fig:random_30} in Appendix). For  random guessing attack, the optimal G-mean value can be achieved by setting the probability to $0.5$. The G-mean for the naive attack is always $0\%$.
\begin{equation}
\footnotesize
{
\mathsf{G-mean = \sqrt{\frac{TP}{TP + FN} * \frac{TN}{TN + FP}}} 
}
\end{equation}
\textbf{Matthews correlation coefficient (MCC)}: This MCC metric also takes into account all of TP, TN, FP, and FN, and is a balanced measure which can be used even if the classes of the sensitive attribute are of very different sizes~\cite{MATTHEWS1975442}. It returns a value between -1 and +1. A coefficient of +1 represents a perfect prediction, 0 represents a prediction no better than the random one, and -1 represents a prediction that is always incorrect. Note that, even if the adversary has the knowledge of marginal priors of the sensitive attribute, it is not able to achieve an MCC value of more than $0$ with the random guessing attack strategy (details in Appendix~\ref{appn:random}). Also, the naive attack always results in an MCC of $0$, independent of the marginal prior knowledge. 
\begin{equation}
\footnotesize
{
\mathsf{MCC = \frac{(TP*TN)-(FP*FN)}{\sqrt{(TP+FP)*(TP+FN)*(TN+FP)*(TN+FN)}}}
}
\end{equation}

\section{New Model Inversion Attacks}
\label{sec:new_attacks}
We design two new attack strategies: 
(1) confidence modeling-based model inversion attack and
(2) confidence score-based model inversion attack.
Table~\ref{table:adversary_assumptions} shows the different adversary capabilities/knowledge assumptions for these attacks in contrast to the existing attacks.

\begin{figure*}[h]
\centering
\includegraphics[width=0.99\textwidth, height=7cm]
{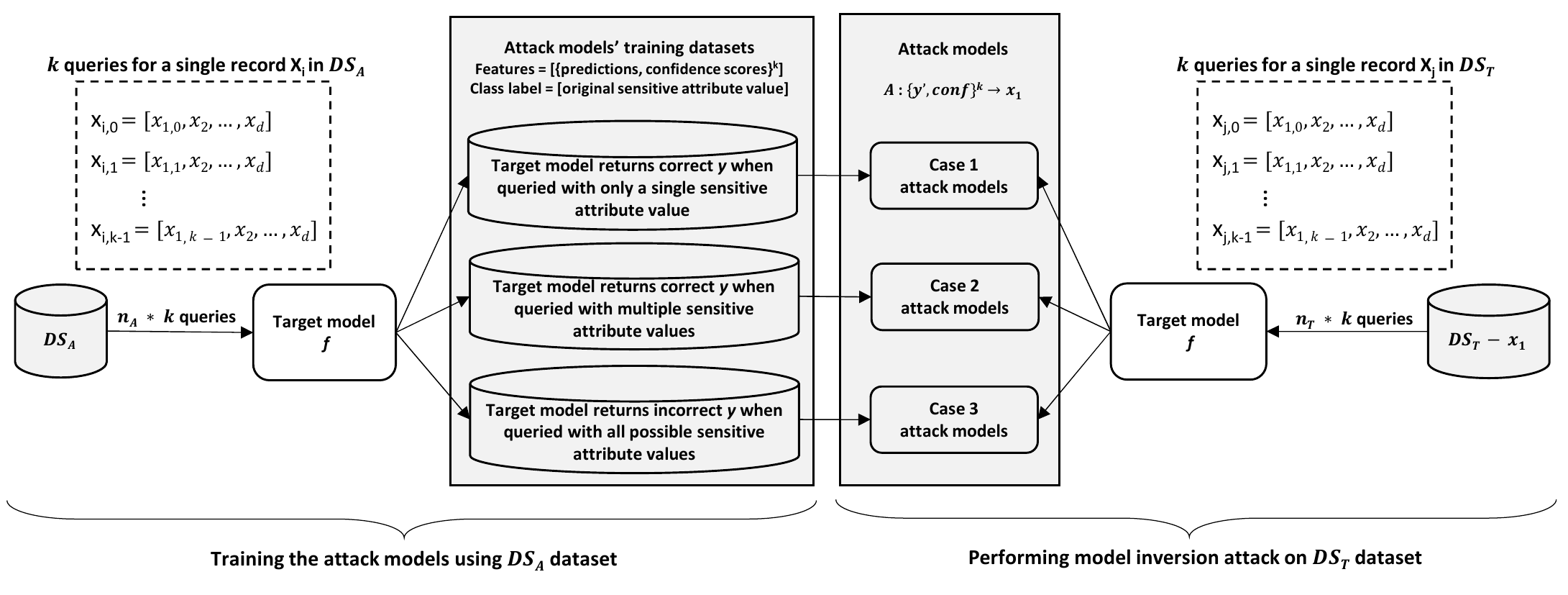}
\caption{Confidence modeling-based model inversion attack (CMMIA). First, the adversary collects the attack models' training datasets by querying the target model $f$ using $DS_A$ dataset to train the attack models. The adversary then leverages the trained attack models to predict the sensitive attribute values of the records in $DS_T$ dataset.}
\label{fig:cmmia}
\end{figure*}

\subsection{Confidence Modeling-based Model Inversion Attack (CMMIA)} 
\label{subsec:cmodel_attack}

In confidence modeling-based model inversion attack, the adversary models the predictions and confidence values returned by the target model $f$ to \emph{predict the sensitive attribute value}. We assume that the adversary has access to a dataset, $DS_A$, collected from the same population the $DS_T$ dataset has been obtained, where $DS_A \cap DS_T = \emptyset$. The adversary first queries the target model $f$ using records in $DS_A$ dataset and collects predictions and confidence scores returned by the target model to train the attack models. The attack steps are described in the following.

\subsubsection{Collecting data for training attack models} 
\label{cmodel_attack_collect}
The adversary first queries the target model $f$ using the records in dataset $DS_A$ (see Fig.~\ref{fig:cmmia}). Note that, the adversary has knowledge of the actual sensitive attributes ($x_1$) in the $DS_A$ dataset along with non-sensitive attributes, $x_2, ... x_d$ and actual $y$. The goal of this attack step is to collect data on how the target model responds to the queries performed with the $DS_A$ dataset, i.e., the predictions and confidence scores returned by the target model, when the adversary varies the sensitive attribute $x_1$.
The number of queries performed for this data collection (attack models' training data) step is: $n_A * k$, where $n_A$ is the number of records in $DS_A$ and $k$ is the number of unique possible values of the sensitive attribute $x_1$. For instance, if there are two possible values of a sensitive attribute (i.e., $k$ = 2, well depicted by a yes/no answer from an individual in response to a survey question), the adversary queries the model by setting the sensitive attribute value $x_1$ to both \emph{yes} and \emph{no} while all other known input attributes of the target individual remain the same. Let $y'_{0}$ and $conf_0$ be the returned model prediction and confidence score when the sensitive attribute is set to  \emph{no}. Similarly, $y'_{1}$ and $conf_1$ are the model prediction and confidence score when the sensitive attribute is set to \emph{yes}. After querying the target model, there are three possible outcomes:
\begin{enumerate}[label=\fbox{\textbf{Case (\arabic*)}},  leftmargin=0pt, itemindent=53pt]
    \item The target model $f$ predicts the correct $y$ \emph{only for a single sensitive attribute value}, e.g., $y$ = $y'_{0}$ $\wedge$ $y$ != $y'_{1}$ or $y$ != $y'_{0}$ $\wedge$ $y$ = $y'_{1}$, in the case of a binary sensitive attribute. The adversary collects the records $[\{y'_0, conf_0\}, ..., \{y'_{k-1}, conf_{k-1}\}, x_1]$, i.e., the predictions with $k$ possible sensitive attribute values and the associated confidence scores, to train the case 1 attack models (see Fig.~\ref{fig:cmmia}). 
    \vspace{0.1cm}
    \item The target model $f$ predicts the correct $y$ for multiple sensitive attribute values, i.e., multiple $y'$s equal $y$. The adversary collects the records $[\forall{i} \{ y'_i, conf_i\}$ \emph{s.t.} $y'_i = y, x_1]$ to train the case 2 attack models. 
    \vspace{0.1cm}
    \item The target model $f$ predicts incorrect $y'$s for all possible sensitive attribute values. The adversary collects the records $[\{y'_0, conf_0\}$ $, ...,$ $\{y'_{k-1}, conf_{k-1}\},$ $x_1]$ to train the case 3 attack models, where for any $i$ in $\{0,..., k-1\}$, $y'_i \neq y$.
\end{enumerate}

\subsubsection{Training the attack models} 
\label{cmodel_attack_train}
The adversary trains $m$ attack models for each of the $3$ cases mentioned above where $m$ is the number of target model class labels (mentioned in Section~\ref{subsec:mia}). The adversary first separates the records under each case into $m$ subsets based on the original target class label ($y$) of the records in $DS_A$. The reason behind training separate attack models for each target model class label is that the target model's performance (returned predictions and confidence scores) can be different for different class labels which may be due to various reasons, e.g., underlying unbalanced training dataset. For example, a vehicle image classification model may perform very good in identifying images of trucks but demonstrate poor performance in identifying images of vans.

In summary, we train $3*m$ attack models that take as input the predictions and confidence scores returned by the target model and outputs a prediction for the sensitive attribute:
\begin{equation}
{
\mathcal{A}_l^q : [\{y'_0, conf_0\}, ..., \{y'_{k-1}, conf_{k-1}\}] \longrightarrow x_1
}
\end{equation}
where $1 \le l \le m$ represents original class label index and $1 \le q \le 3$ represents case number. For instance, $\mathcal{A}_2^3$ represents the attack model that is trained to predict the sensitive attribute $x_1$ for a record when the original class label of that record is $y_2$ and the target model returns incorrect predictions for all $k$ queries, i.e., for any $i$ in $\{0,..., k-1\}$, $y'_i \neq y_2$.
Note that, in case (2), the input of attack models can have less features since the adversary collects the target model's right predictions only, $[\forall{i} \{ y'_i, conf_i\}$ \emph{s.t.} $y'_i = y, x_1]$, as described in Section~\ref{cmodel_attack_collect}.

\subsubsection{Performing model inversion attack on $DS_T$} 
\label{cmodel_attack_attack}
The adversary now performs $n_T * k$ queries to the target model $f$, where $n_T$ is the number of records in $DS_T$ and $k$ is the number of unique possible values of the sensitive attribute $x_1$. Unlike $DS_A$, the adversary does not know the $x_1$ attribute of the records in $DS_T$ and the goal here is to estimate this attribute. 

\textit{When querying the target model $f$ with $DS_T$ dataset, the adversary classifies the outcomes into three cases in the same way as described in Section~\ref{cmodel_attack_collect}.} Whereas in Section~\ref{cmodel_attack_collect} the adversary collects data for training the attack models, here the adversary leverages the trained attack models to estimate the sensitive attribute $x_1$. For instance, to estimate the sensitive attribute $x_1$ of a record with \emph{original class label $y_2$}, the adversary first queries the target model $f$ with varying sensitive attribute values and obtains the corresponding predictions and confidence score pairs $\{y'_0, conf_0\}, ..., \{y'_{k-1}, conf_{k-1}\}$. The adversary then inputs these values to an attack model based on the outcomes of the target model. If the outcomes of the target model fall in--
\begin{itemize}
    \item case (1), the adversary queries attack model $\mathcal{A}_2^1$. 
    \item case (2), the adversary queries attack model $\mathcal{A}_2^2$. 
    \item case (3), the adversary queries attack model $\mathcal{A}_2^3$. 
\end{itemize}
Finally, the attack model outputs a prediction for the sensitive attribute $x_1$.

\subsection{Confidence Score-based Model Inversion Attack (CSMIA)} 
\label{subsec:cscore_attack}
We propose another attack that exploits the confidence scores returned by the model but in this attack the adversary does not have access to the dataset $DS_A$. Therefore, this attack is suitable for an adversary that is not able to obtain such a dataset.
Unlike Fredrikson et al.~\cite{FredriksonCCS2015} attack, the adversary assumed in this attack does not have access to the marginal priors or the confusion matrix. Similar to the steps described in Section~\ref{subsec:cmodel_attack}, the adversary queries the model by setting the sensitive attribute value $x_1$ to all possible $k$ values while all other known input attributes of the target individual remain the same. \emph{The key idea of this attack is that the target model's returned prediction is more likely to be correct and the confidence score is more likely to be higher when it is queried with a record containing the real sensitive attribute value (since the target model encountered the record in the training dataset with the real sensitive attribute value)}. In order to determine the value of $x_1$, this attack considers the following cases:
    \begin{enumerate}[label=\fbox{\textbf{Case (\arabic*)}},  leftmargin=0pt, itemindent=53pt]
        \item If the target model predicts the correct $y$ \emph{only for a single sensitive attribute value}, e.g., $y$ = $y'_{0}$ $\wedge$ $y$ != $y'_{1}$ or $y$ != $y'_{0}$ $\wedge$ $y$ = $y'_{1}$, in the case of a binary sensitive attribute, the attack selects the sensitive attribute to be the one for which the prediction $y'$ matches $y$. For instance, if $y$ = $y'_{1}$ $\wedge$ $y$ != $y'_{0}$, the attack predicts $yes$ for the sensitive attribute and vice versa. 
        \vspace{0.3cm}
        \item If the model predicts the correct $y$ for multiple sensitive attribute values, i.e., $y$ = $y'_{0}$ $\wedge$ $y$ = $y'_{1}$, the attack selects the sensitive attribute to be the one for which the prediction confidence score is the maximum. In the above example, if the model predicts the $y$ value with a higher confidence when $yes$ value is set for the sensitive attribute, the attack outputs the $yes$ value for the $x_1$ prediction and vice versa. 
        \vspace{0.1cm}
        \item If the model outputs incorrect predictions for all possible sensitive attribute values, i.e., $y$ != $y'_{0}$ $\wedge$ $y$ != $y'_{1}$, the attack selects the sensitive attribute to be the one for which the prediction confidence is the minimum. In the above example, if the model outputs the incorrect prediction with a higher confidence when $yes$ value is set for the sensitive attribute, the attack outputs the $no$ value for the $x_1$ prediction and vice versa. 
    \end{enumerate}

\section{Attack With Partial Knowledge of Target Individual's Non-sensitive Attributes}
\label{sec:new_attacks_partial}
All of our proposed attacks mentioned in Section~\ref{sec:new_attacks} as well as the Fredrikson et al.~\cite{FredriksonCCS2015} attack assume that the adversary has full knowledge of the target individual's non-sensitive attributes. Although these attacks raise serious privacy concerns against a model trained on sensitive dataset, it is not clear how much risk is incurred by these model inversion attacks if the adversary has only partial access to the other (non-sensitive) attributes. In many cases, it may be difficult or even impossible for an adversary to obtain all of the non-sensitive attributes of a target individual. Therefore, the goal of this section is to quantify the risk of model inversion attacks in the case where all non-sensitive attributes of a target individual are not available to the adversary.

In the following, we describe our \emph{confidence score-based model inversion attack} for this special case. Therefore, this model inversion attack with partial knowledge of target individual's non-sensitive attributes does not require the adversary to have access to the dataset $DS_A$.

For simplicity, we assume that there is only one non-sensitive attribute that is unknown to the adversary. Extending our attack steps to more than one unknown attribute is straightforward.
In Section~\ref{subsec:expres_mia_partial}, we conduct experiments when a single as well as multiple non-sensitive attributes are unknown to the adversary.
Without loss of generality, let $x_2 \in \mathbf{x}$ be the non-sensitive attribute unknown to the adversary. 
Also, let $u$ be the number of unique possible values of $x_2$. We query the model by varying the unknown non-sensitive attribute with its different unique possible values (in the same way we vary the sensitive attribute $x_1$ in the attacks described in Section~\ref{sec:new_attacks}) while all other known non-sensitive attributes $\{x_3, ..., x_d\}$ remain the same. Hence, in this attack, we query the model $u$ times for each possible value of the sensitive attribute. As a result, the complexity of the attacks described in this section is $u$ times the complexity of the attacks in Section~\ref{sec:new_attacks}. 

According to the notations used in Section~\ref{sec:new_attacks}, let $C_{0}$=${\textstyle\sum}_{i=1}^{u} (y = y'_{0\_i})$ be the number of times the predictions are correct with the sensitive attribute \emph{no} and $C_{1}$=${\textstyle\sum}_{i=1}^{u} (y = y'_{1\_i})$  be the number of times the predictions are correct with the sensitive attribute \emph{yes}.

In order to determine the value of $x_1$, this attack considers the following cases:
    \begin{enumerate}[label=\fbox{\textbf{Case (\arabic*)}},   leftmargin=0pt, itemindent=53pt]
        \item If $C_{0}$ != $C_{1}$, i.e., the number of correct target model predictions are different for different sensitive attribute values, the attack selects the sensitive attribute to be the one for which the number of correct predictions is higher. For instance, if $C_{1}>$  $C_{0}$, the attack predicts $yes$ for the sensitive attribute and vice versa. 
        \vspace{0.1cm}
        \item If $C_{0}$ = $C_{1}$ and both are non-zero, we compute the sum of the confidence scores (only for the correct predictions) for each sensitive attribute and the attack selects the sensitive attribute to be the one for which the sum of the confidence scores is the maximum. 
        \vspace{0.1cm}
        \item If $C_{0} = 0$ $\wedge$ $C_{1} = 0 $, we compute the sum of the confidence scores for each sensitive attribute and the attack selects the sensitive attribute to be the one for which the sum of the confidence scores is the minimum.
    \end{enumerate}
If there is a second non-sensitive attribute that is unknown to the adversary (let that unknown attribute be $x_3$) and $v$ is the number of unique possible values for that unknown non-sensitive attribute, we query the model by varying both $x_2$ and $x_3$ while all other known non-sensitive attributes $\{x_4, ..., x_d\}$ remain the same. Hence, in this attack, we query the model $u * v$ times for each possible value of the sensitive attribute. As a result, the complexity of the attack becomes $u * v$ times the complexity of the attacks in Section~\ref{sec:new_attacks}.

\section{Evaluations}
\label{sec:expres}
In this section, we discuss our experiment setup (i.e., datasets, machine learning models, and performance metrics) and evaluate our proposed attacks. \emph{To facilitate reproducibility, we will publicly release the code-base, all the datasets, and machine learning models used in the following experiments.}

\subsection{Datasets}
\label{subsec:expres_data}
\textbf{General Social Survey (GSS)~\cite{gss}}:
Fredrikson et al.~\cite{FredriksonCCS2015} attack, denoted as \emph{FJR} attack in the rest of this paper, uses the \emph{General Social Survey (GSS)} dataset to demonstrate their attack effectiveness. This dataset has $51,020$ records with $11$ attributes and is used to train a model that predicts how happy an individual is in his/her marriage. However, the training dataset for this model contains sensitive attribute about the individuals: e.g., responses to the question `\emph{Have you watched X-rated movies in the last year?}'. Removing the data records that do not have either the sensitive attribute or the attribute that is being predicted by the target model (i.e., happiness in marriage) results in $20,314$ records that we use in our experiments. For CMMIA, $DS_A$ consists of randomly chosen $5,079$ records and the rest $15,235$ records belong to $DS_T$ (i.e., a 25\%-75\% split). To ensure consistency, we evaluate our CSMIA attack and other baseline attack strategies including FJR attack~\cite{FredriksonCCS2015} on the target models trained on the $DS_T$ dataset ($15,235$ records). Among the $15,235$ records in the $DS_T$ dataset, $3,017$ individuals answered \emph{yes} (i.e., sensitive attribute $x_1$ = \emph{yes}) to the survey question on whether they watched X-rated movies in the last year and the rest $12218$ individuals answered \emph{no} (i.e., $x_1$ = \emph{no}). 

 \begin{table}[t]
  \centering
  \caption{Distribution of sensitive attributes in  datasets.}
  \resizebox{1\columnwidth}{!}{
  \begin{tabular}{ | l | l | l | l | c | l |}
    \hline
    \multirow{2}{*}{Dataset}  & \multirow{2}{*}{Sensitive attribute}  &  Positive   & Negative   & Positive   & Positive   \\
            &      &  class label  & class label & class count   & class \%  \\ \hline
    \multirow{2}{*}{GSS}     & \multirow{2}{*}{Watched x-rated movie}    & \multirow{2}{*}{Yes} & \multirow{2}{*}{No} & 4002   & 19.70\% \\
            &    &               &             &    (3017)      &     (19.80\%)     \\ \hline

    \multirow{2}{*}{Adult}  & \multirow{2}{*}{Marital status} & \multirow{2}{*}{Married} & \multirow{2}{*}{Single} & 21639    & 47.85\%  \\
            &         &               &             &  (16893)        &   (47.96\%)  \\ \hline
  \end{tabular}
  }
\label{table:data_distribution}
\end{table}

\textbf{Adult Dataset~\cite{adult}}:
This dataset, also known as \emph{Census Income} dataset, is used to predict whether an individual earns over \$50K a year. The number of instances in this dataset is $48,842$ and it has $14$ attributes. We merge the `marital status' attribute into two distinct clusters, Married: \{Married-civ-spouse, Married-spouse-absent, Married-AF-spouse\} and Single: \{Divorced, Never-married, Separated, Widowed\}. We then consider this attribute (Married/Single) as the sensitive attribute that the adversary aims to learn.  After removing the data records with missing values, the final dataset consists of $45222$ records. For CMMIA, $DS_A$ consists of randomly chosen $10,000$ records and the rest $35,222$ records belong to $DS_T$. To ensure consistency, we evaluate all the model inversion attacks in comparison (proposed and baseline) against the target models trained on the $DS_T$ dataset ($35,222$ records). Among the $35222$ records, $16893$ individuals are \emph{married} (i.e., sensitive attribute $x_1$ = \emph{married}) and the rest $18329$ individuals are \emph{single} (i.e., $x_1$ = \emph{single}). For the experiments in this paper, we have removed the `\emph{relationship}' attribute from this dataset since this attribute's values (e.g., husband, wife, unmarried) are directly related to the \emph{marital status} attribute that the model inversion attacks aim to learn.

\begin{table*}[h]
  \centering
  \caption{Attack performance against the decision tree target model trained on GSS dataset.}
  \resizebox{0.85\textwidth}{!}{
  \begin{tabular}{ | l | l | l | l | l | l | l | l | l | l | l |}
    \hline
    Attack Strategy &  TP & TN & FP & FN & Precision & Recall & Accuracy & F1 score & G-mean & MCC \\ \hline 
    Naive attack & $0$ & $12218$ & $0$ & $3017$ & $0\%$ & $0\%$ & $80.2\%$ & $0\%$ & $0\%$ & $0\%$\\ \hline 
    FJR attack~\cite{FredriksonCCS2015} & $131$ & $11709$ & $509$ & $2886$ & $20.47\%$ & $4.34\%$ & $77.72\%$ & $7.16\%$ & $20.39\%$ & $0.3\%$\\ \hline 
    Confidence modeling-based attack & $1766$ & $7605$ & $4610$ & $1254$ & $27.7\%$ & $58.48\%$ & $61.51\%$ & $37.59\%$ & $60.34\%$ & $16.8\%$\\ \hline 
    Confidence score-based attack & $1490$ & $7844$ & $4373$ & $1528$ & $25.41\%$ & $49.37\%$ & $61.27\%$ & $33.55\%$ & $56.3\%$ & $11.1\%$\\ \hline 
    
  \end{tabular}
  }

\label{table:gss_results_dt}

\vspace{0.3cm}

  \centering
  \caption{Attack performance against the deep neural network target model trained on GSS dataset.}
  \resizebox{.85\textwidth}{!}{
  \begin{tabular}{ | l | l | l | l | l | l | l | l | l | l | l |}
    \hline
    Attack Strategy &  TP & TN & FP & FN & Precision & Recall & Accuracy & F1 score & G-mean & MCC \\ \hline 
    FJR attack~\cite{FredriksonCCS2015} & $1$ & $12213$ & $5$ & $3016$ & $16.67\%$ & $0.03\%$ & $80.17\%$ & $0.07\%$ & $1.82\%$ & $-0.2\%$\\ \hline 
    Confidence modeling-based attack & $1100$ & $8133$ & $4085$ & $1917$ & $21.22\%$ & $36.46\%$ & $60.61\%$ & $26.82\%$ & $49.26\%$ & $2.2\%$\\ \hline 
    Confidence score-based attack & $1212$ & $8058$ & $4160$ & $1805$ & $22.56\%$ & $40.17\%$ & $60.85\%$ & $28.89\%$ & $51.47\%$ & $5.1\%$\\ \hline

  \end{tabular}
  }
\label{table:gss_results_dnn}
\end{table*}

\subsection{Machine Learning Models}
\label{subsec:expres_model}
In order to evaluate our proposed attacks and other existing attacks, we train target models on each of the two datasets mentioned in Section~\ref{subsec:expres_data} using two different machine learning techniques, decision tree and deep neural network. The confusion matrices of all the trained models are given in Appendix (Tables~\ref{table:CM_DT_GSS},~\ref{table:CM_DNN_GSS},~\ref{table:CM_DT_Adult}, and~\ref{table:CM_DNN_Adult}). We leverage BigML~\cite{bigml}, an ML-as-a-service system, to train these target models. Users can leverage such a service by uploading their datasets, selecting an attribute as the objective field, and training a model to predict that objective field when the other attributes are given as input. BigML allows users to publish their models in black-box mode, i.e., the models can be queried by other users and they can obtain the model predictions along with the confidence scores. CMMIA strategy's attack models are also trained using BigML's decision tree machine learning technique.

\subsection{Attack Performance Metrics}
\label{subsec:expres_metric}
As mentioned earlier, the \emph{accuracy} metric may fail to evaluate an attack or even misrepresent the attack performance if the dataset is unbalanced. Table~\ref{table:data_distribution} shows the distribution of sensitive attribute values in the GSS and Adult datasets (positive class count and \% in $DS_T$ are shown in parenthesis). Since the sensitive attribute in the GSS dataset is unbalanced, a naive attack always predicting the negative class would result in $\sim80\%$ accuracy, which is a misleading evaluation of attack performance. Moreover, the \emph{F1 score} alone is not a meaningful metric to evaluate the attacks since it emphasizes only on the positive class. Therefore, we also use \emph{G-mean} and \emph{MCC} metrics as described in Section~\ref{sec:indepth} to evaluate our attacks as well as to compare their performances with that of the FJR attack~\cite{FredriksonCCS2015} and the baseline attacks (naive and random guessing).

\subsection{New Model Inversion Attacks' Results}
\label{subsec:expres_mia}

\subsubsection{GSS Dataset}
\label{subsubsec:expres_mia_gss}
Tables~\ref{table:gss_results_dt} and~\ref{table:gss_results_dnn} show the performance of the baseline attacks and our proposed attacks against the decision tree and deep neural network target models trained on the GSS dataset, respectively. 
As shown in the results, the FJR attack~\cite{FredriksonCCS2015} achieves a very low recall and thus low F1 score. This is due to the fact that the FJR attack~\cite{FredriksonCCS2015} relies on the marginal prior of the sensitive attribute while performing the attack (described in Section~\ref{subsec:ccs15}). Since the sensitive attribute in the GSS dataset is unbalanced, the FJR attack~\cite{FredriksonCCS2015} mostly predicts the negative sensitive attribute (i.e., the individual didn't watch x-rated movie, marginal prior $\sim0.8$) and rarely predicts the positive sensitive attribute (i.e., the individual watched x-rated movie, marginal prior $\sim0.2$). The FJR attack~\cite{FredriksonCCS2015} against the deep neural network target model  performs almost like a naive attack with only 1 true positive and 5 false positives as demonstrated in Table~\ref{table:gss_results_dnn}. 
In contrast, our proposed CMMIA and CSMIA strategies achieve significantly high recall, F1 score, G-mean, and MCC while also improving precision. The FJR attack~\cite{FredriksonCCS2015} performs better only in terms of accuracy. However, note that the naive attack also achieves an accuracy of $80.2\%$, the highest among all attacks, but there is no attack efficacy (0 true positive).

In CMMIA, for each of the three cases, we train three decision tree models, one for each possible $y$ value of the target model (i.e., happiness in marriage-- `not too happy', `pretty happy', and `very happy'), i.e., nine decision tree models in total. In most of the experiments, the CMMIA strategy performs better than CSMIA since it leverages the attack models. Note that, attack models take the predictions and confidence scores returned by the target model when queried with varying sensitive attributes as input and outputs a prediction for the sensitive attribute. As a result, for instance, in case (2), the CMMIA strategy learns that even if the target model returns higher confidence for the negative sensitive attribute (i.e., the individual didn't watch x-rated movie), the actual sensitive attribute value could be positive (i.e., the individual watched x-rated movie) and thus reports more true positives (see case (2) results in Tables~\ref{table:gss_dt_bpmpcs_details} and~\ref{table:gss_dnn_bpmpcs_details} in Appendix that show the contrast between the confidence modeling-based and confidence score-based attacks in details).

Note that, all our attacks perform consistently across different machine learning models. In contrast, the FJR attack~\cite{FredriksonCCS2015} shows notably different performance in terms of identifying the positive cases against the deepnet target model.

\begin{table*}[t]
  \centering
  \caption{Attack performance against the decision tree target model trained on  Adult dataset.}
  \resizebox{.85\textwidth}{!}{
  \begin{tabular}{ | l | l | l | l | l | l | l | l | l | l | l |}
    \hline
    Attack Strategy &  TP & TN & FP & FN & Precision & Recall & Accuracy & F1 score & G-mean & MCC \\ \hline
    Naive attack & $0$ & $18329$ & $0$ & $16893$ & $0\%$ & $0\%$ & $52.04\%$ & $0\%$ & $0\%$ & $0\%$\\ \hline 
    FJR attack~\cite{FredriksonCCS2015} & $3788$ & $17818$ & $511$ & $13105$ & $88.11\%$ & $22.42\%$ & $61.34\%$ & $35.75\%$ & $46.69\%$ & $29.9\%$\\ \hline 
    Confidence modeling-based attack & $12311$ & $11619$ & $6710$ & $4582$ & $64.72\%$ & $72.88\%$ & $67.94\%$ & $68.56\%$ & $67.97\%$ & $36.4\%$\\ \hline 
    Confidence score-based attack & $7664$ & $17085$ & $1244$ & $9229$ & $86.04\%$ & $45.37\%$ & $70.27\%$ & $59.41\%$ & $65.03\%$ & $44.3\%$\\ \hline

  \end{tabular}
  }
\label{table:adult_results_dt}

\vspace{0.3cm}

  \centering
  \caption{Attack performance against the deep neural network target model trained on Adult dataset.}
  \resizebox{.85\textwidth}{!}{
  \begin{tabular}{ | l | l | l | l | l | l | l | l | l | l | l |}
    \hline
    Attack Strategy &  TP & TN & FP & FN & Precision & Recall & Accuracy & F1 score & G-mean & MCC \\ \hline
    FJR attack~\cite{FredriksonCCS2015} & $3592$ & $17717$ & $612$ & $13301$ & $85.44\%$ & $21.26\%$ & $60.5\%$ & $34.05\%$ & $45.34\%$ & $27.62\%$\\ \hline 
    Confidence modeling-based attack & $11907$ & $11328$ & $7001$ & $4986$ & $62.97\%$ & $70.48\%$ & $65.97\%$ & $66.52\%$ & $66.01\%$ & $32.35\%$\\ \hline 
    Confidence score-based attack & $7490$ & $17139$ & $1190$ & $9403$ & $86.29\%$ & $44.34\%$ & $69.93\%$ & $58.58\%$ & $64.39\%$ & $43.87\%$\\ \hline

  \end{tabular}
  }
\label{table:adult_results_dnn}
\end{table*}

\subsubsection{Adult Dataset}
\label{subsubsec:expres_mia_adult}
Tables~\ref{table:adult_results_dt} and~\ref{table:adult_results_dnn} show the performance of the baseline attacks and our proposed attacks against the decision tree and deep neural network target models trained on the Adult dataset, respectively. While our CMMIA strategy results in the highest recall, the CSMIA attack performs better in terms of precision. For the CMMIA strategy, for each of the three cases, we train two decision tree models, one for each possible $y$ value of the target model (i.e., income-- `<=50K' and `>50K'), i.e., six decision tree models in total. 
Tables~\ref{table:adult_dt_bpmpcs_details} and~\ref{table:adult_dnn_bpmpcs_details} 
in Appendix show the contrast between CMMIA and CSMIA in details.

Overall, the attacks against the target models trained on Adult dataset demonstrate more effectiveness than that of against the target models trained on GSS dataset. However, the correlations between the sensitive attributes and the corresponding target models trained on these datasets (in other words, importance of the sensitive attributes in the target models) do not differ significantly. For instance, the importance of the `x-rated-movie' and `marital-status' sensitive attributes in their corresponding decision tree target models are $7.3\%$ and $9.6\%$, respectively. Fig.~\ref{fig:importance} in Appendix shows the importance of all attributes in these models.

\subsection{Understanding Model Inversion Attacks In-depth}
\label{subsec:expres_indepth}
As described in Section~\ref{sec:indepth}, the goal of the following experiments is to understand whether releasing the black-box model really adds more advantage to the adversary to learn the sensitive attributes in the training dataset. 
Therefore, we compare all model inversion attacks (both existing and our proposed ones), i.e., Fredrikson et al. attack (FJRMIA~\cite{FredriksonCCS2015}), confidence-modeling based attack (CMMIA), and confidence score-based attack (CSMIA) with baseline attack strategies that do not require access to the target model, i.e., naive attack (NaiveA) and random guessing attack (RandGA).  Note that, the case (1) of CSMIA does not even require the knowledge of confidence scores. The adversary can perform the case (1) CSMIA attack with the knowledge of the predicted labels ($y$) only. Therefore, we pay special attention to the case (1) of CSMIA and analyze its performance along with the other attacks.

In random guessing attack (RandGA), always predicting the positive class would result in a $100\%$ recall and thus a high F1 score but a G-mean of $0\%$. Therefore, for all the comparisons in the following, the RandGA strategy predicts the positive class with a probability of 0.5, thus maximizing G-mean at $50\%$ and also ensuring a recall of $50\%$. Figures~\ref{fig:random_gss} and~\ref{fig:random_adult} in Appendix show the optimal performance of random guessing attack on GSS and Adult datasets, respectively.

\begin{figure*}[t]
\centering
\subfigure[Decision tree model trained on GSS dataset]{\includegraphics[width=0.47\textwidth, height=4.1cm]
{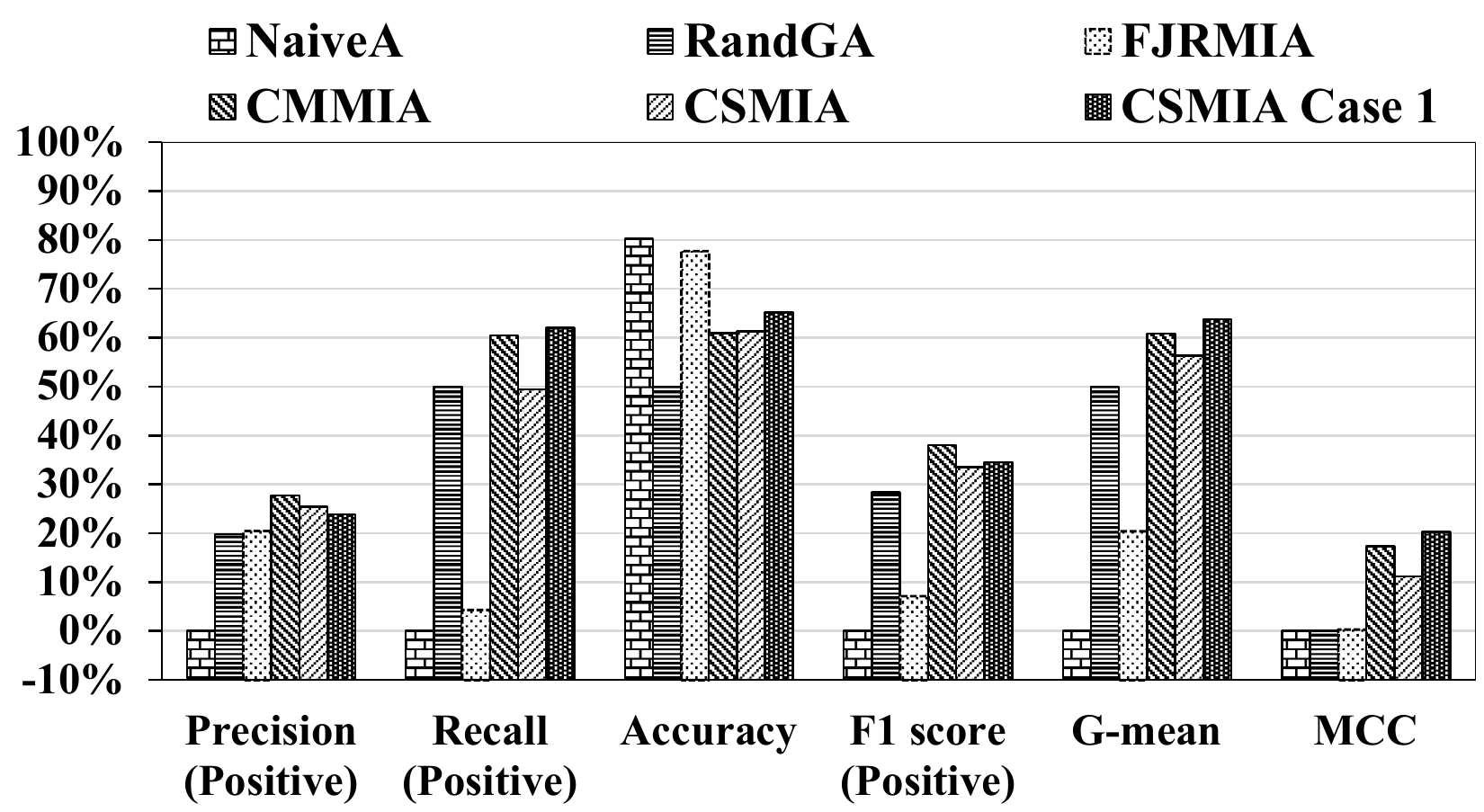}
\label{fig:GSS_RandomDT_NEW}
}
\subfigure[Deepnet model trained on GSS dataset]{\includegraphics[width=0.47\textwidth, height=4.1cm]
{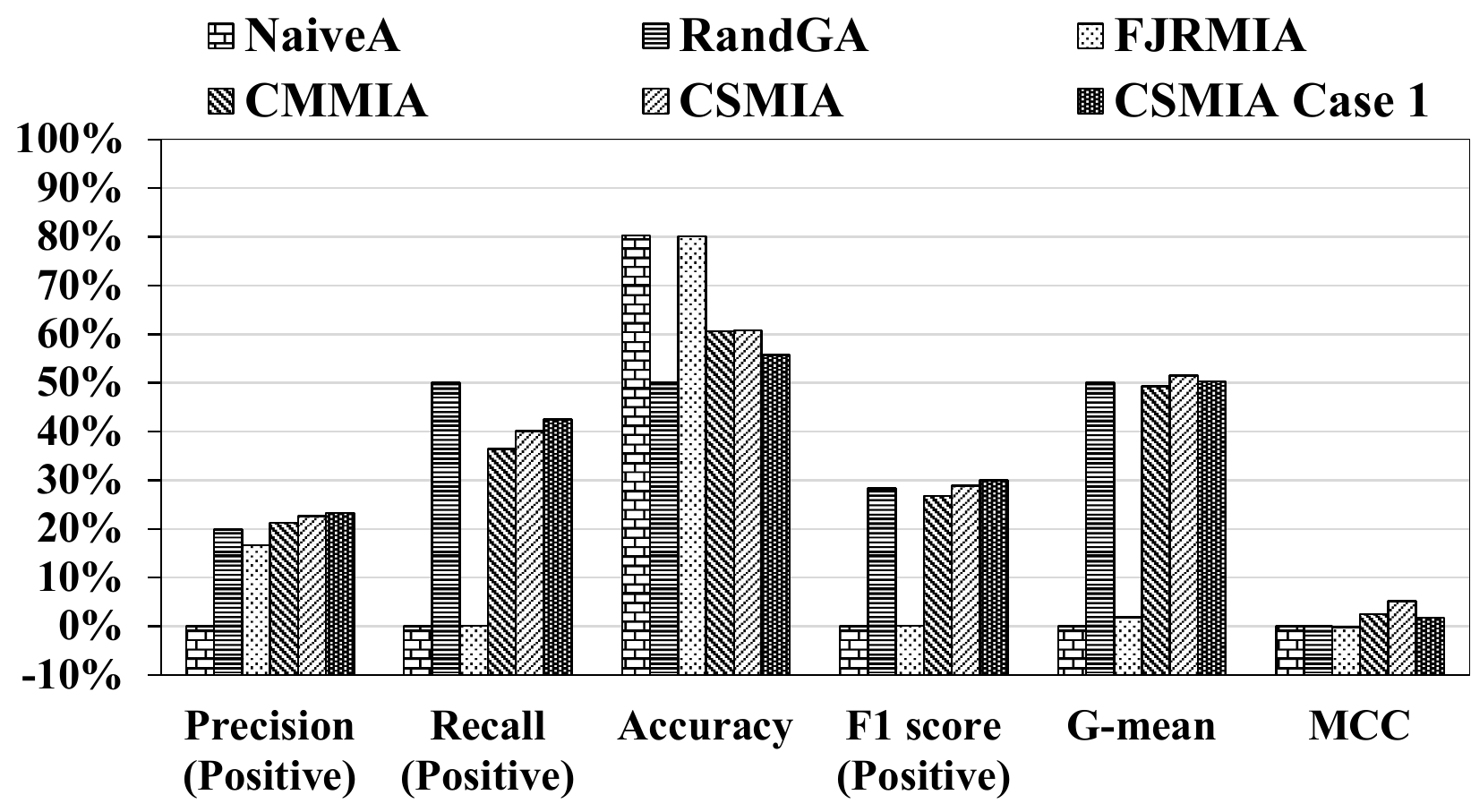}
\label{fig:GSS_RandomDNN_NEW}
}
\subfigure[Decision tree model trained on Adult dataset]{\includegraphics[width=0.47\textwidth, height=4.1cm]
{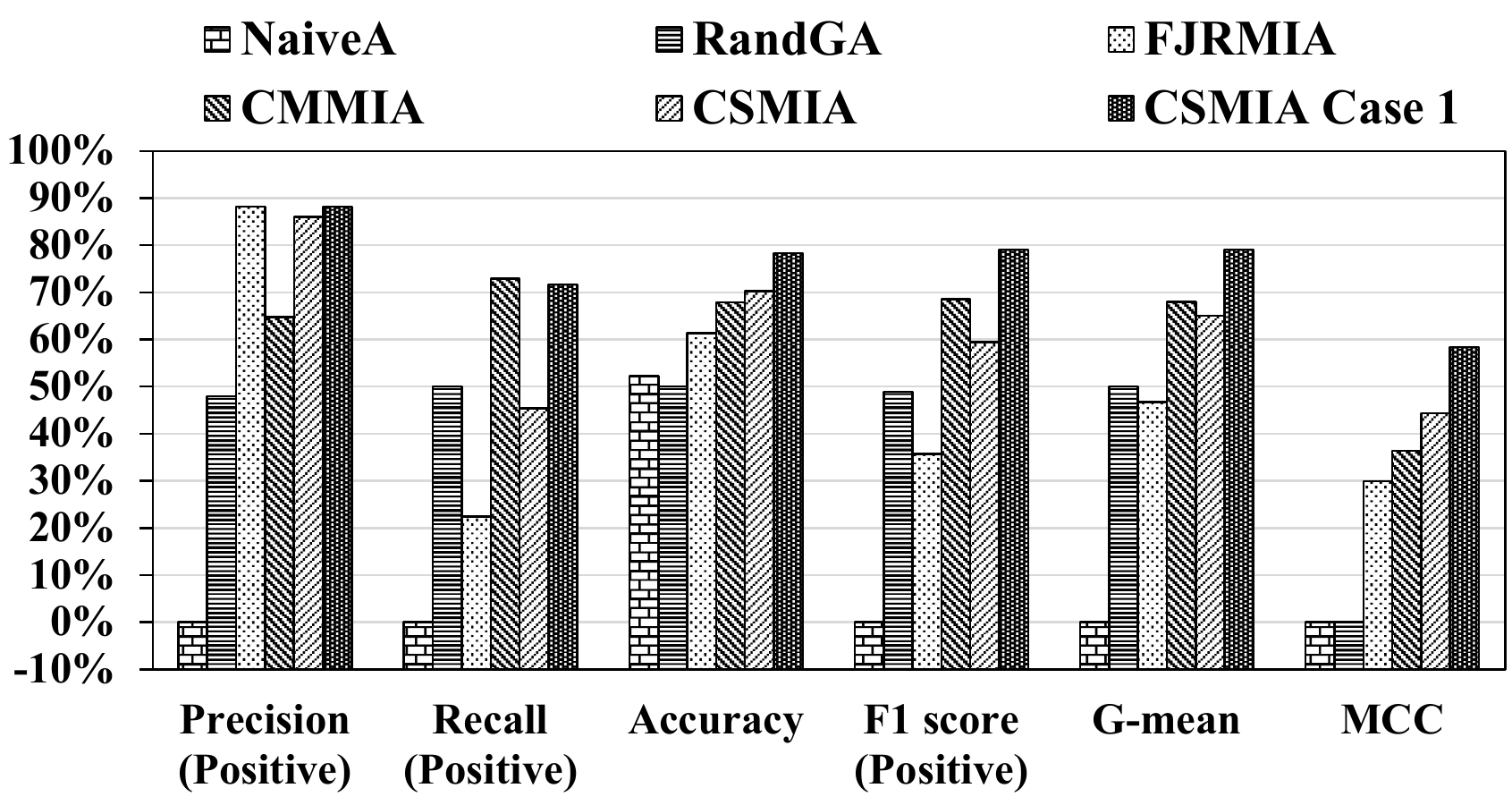}
\label{fig:Adult_RandomDT_NEW}
}
\subfigure[Deepnet model trained on Adult dataset]{\includegraphics[width=0.47\textwidth, height=4.1cm]
{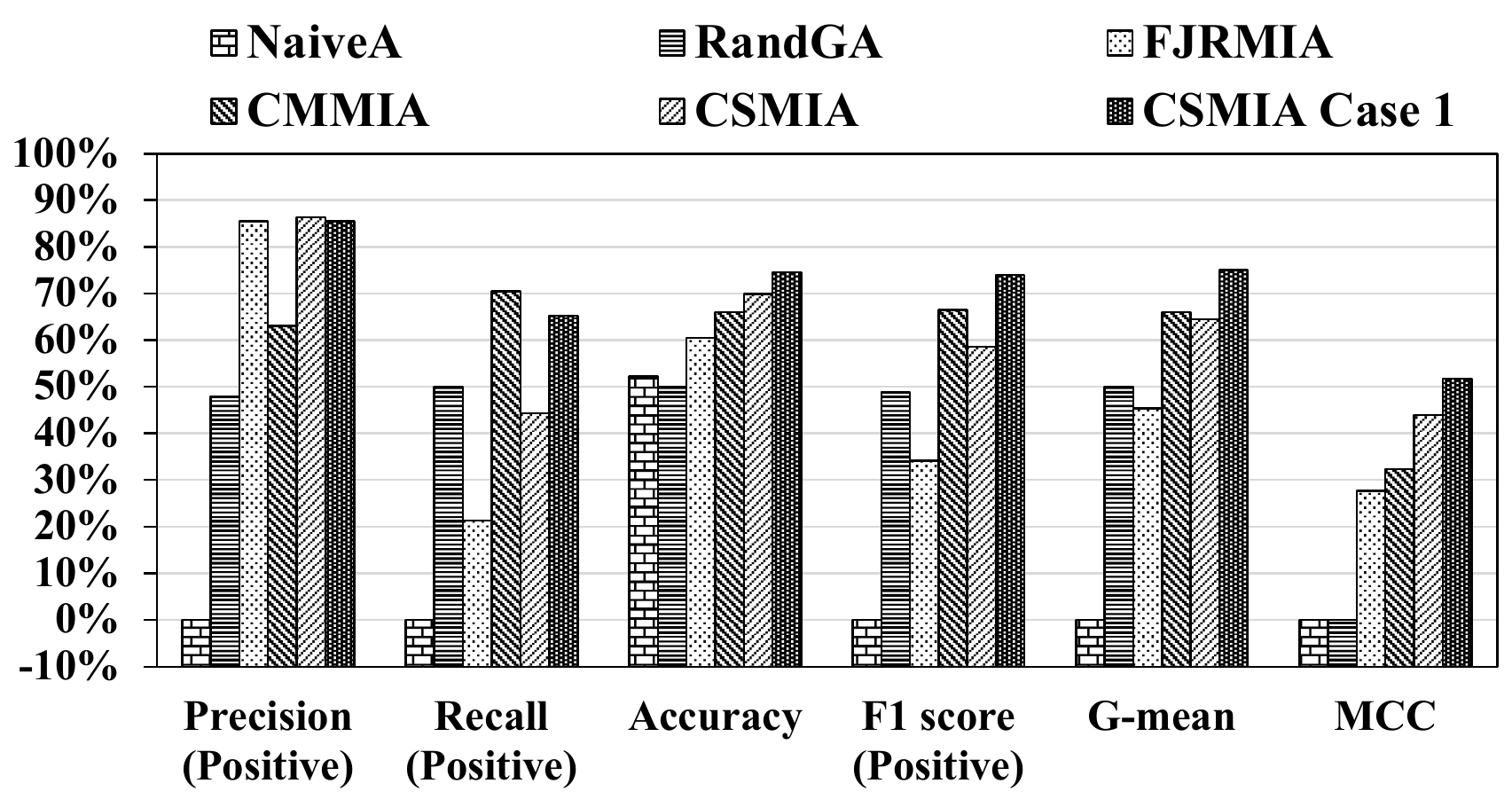}
\label{fig:Adult_RandomDNN_NEW}
}
\vspace{-0.3cm}
\caption{Comparison of model inversion attacks: Fredrikson et al. attack (FJRMIA~\cite{FredriksonCCS2015}), confidence-modeling based attack (CMMIA), confidence score-based attack (CSMIA), and case (1) of CSMIA with baseline attack strategies: naive attack (NaiveA) and random guessing attack (RandGA).}
\label{fig:random}
\end{figure*}

\subsubsection{GSS Dataset}
\label{subsubsec:expres_indepth_gss}
Since the sensitive attribute in this dataset has an unbalanced distribution, the NaiveA strategy (described in Section~\ref{subsec:existing_attacks}), also mentioned in~\cite{FredriksonCCS2015}, predicts the sensitive attribute as \emph{no} for all the individuals and achieves an accuracy of $80.3\%$. However, the precision, recall, F1 score, G-mean, and MCC would all be $0\%$ as shown in Figures~\ref{fig:GSS_RandomDT_NEW} and~\ref{fig:GSS_RandomDNN_NEW}. 

Fig.~\ref{fig:GSS_RandomDT_NEW} shows how the existing and our proposed model inversion attacks compare with NaiveA and RandGA against the decision tree model trained on the GSS dataset. As demonstrated in the figure, the FJRMIA~\cite{FredriksonCCS2015} strategy achieves high accuracy, similar to NaiveA, but does not perform well in terms of any other metrics. Our attacks consistently perform better than RandGA in terms of all metrics. We emphasize that the individuals who belong to the case (1) are more vulnerable to CSMIA. 

Fig.~\ref{fig:GSS_RandomDNN_NEW} shows how the existing and our proposed model inversion attacks compare with NaiveA and RandGA against the deep neural network model trained on the GSS dataset. As demonstrated in the figure, the FJRMIA~\cite{FredriksonCCS2015} strategy achieves a high accuracy but an extremely low recall. The RandGA strategy has the same results as Fig.~\ref{fig:GSS_RandomDT_NEW} since this strategy is independent of the machine learning model. Our attacks' performances against this model are not significantly better than RandGA, even the CSMIA case (1) does not show significant results. Therefore, it may seem that according to the overall performance, the deep neural network model trained on the GSS dataset may not be vulnerable to model inversion attacks since another adversary even without access to the model may achieve comparable performances with RandGA. \emph{However, it is very important to note that the RandGA strategy predicts the sensitive attribute randomly whereas the model inversion attacks rely on the outputs of a model that is trained on the dataset containing the actual sensitive attributes. Even if the overall performance of a model inversion attack on the entire dataset does not seem to be a threat, some specific groups of individuals in the dataset could still be vulnerable.} We discuss such discrimination in performances of model inversion attacks later in Section~\ref{subsec:expres_target}.

\subsubsection{Adult Dataset}
\label{subsubsec:expres_indepth_adult}
Since the sensitive attribute is more balanced in this dataset, the NaiveA strategy has an accuracy of only $52.1\%$, and the other metrics are at $0\%$, similar to that of previous results of this attack. 

As demonstrated in Fig.~\ref{fig:Adult_RandomDT_NEW}, FJRMIA~\cite{FredriksonCCS2015} results in a precision comparable to our attacks but achieves much less in terms of the other metrics. Our attacks significantly outperform RandGA in terms of all metrics except the recall of CSMIA. Fig.~\ref{fig:Adult_RandomDNN_NEW} shows results very similar to that of Fig.~\ref{fig:Adult_RandomDT_NEW}. Observing the results of our proposed attacks and also the CSMIA case (1) performance, we conclude that \emph{releasing the models trained on the Adult dataset} would add significant advantage to the adversary in terms of learning the `marital status' sensitive attribute. This is because all our proposed attacks that query the models for sensitive attribute inference perform significantly better when compared to the RandGA adversary that does not need access to the model.

\begin{figure*}[t]
\centering
\subfigure[CSMIA on decision tree model]{\includegraphics[width=0.48\columnwidth, height=4cm]
{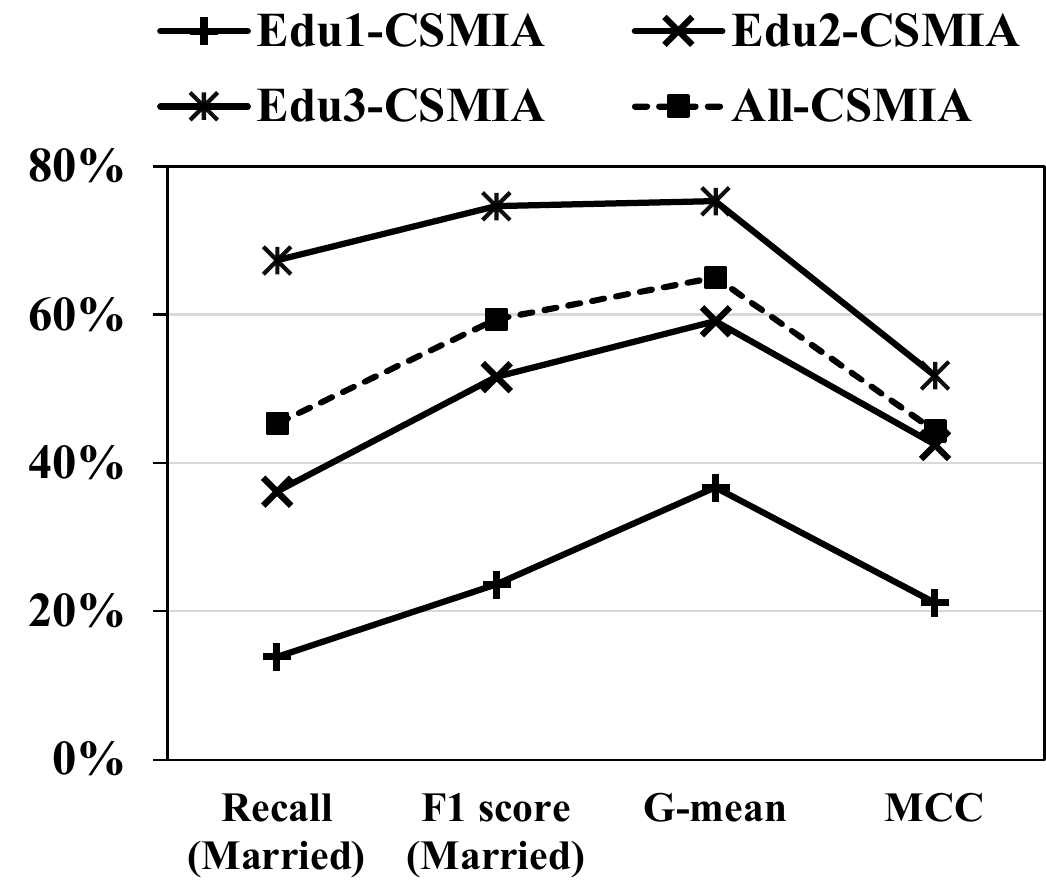}
\label{fig:edu_dt_csmia}
}
\hfill
\subfigure[CSMIA on deepnet model]{\includegraphics[width=0.48\columnwidth, height=4cm]
{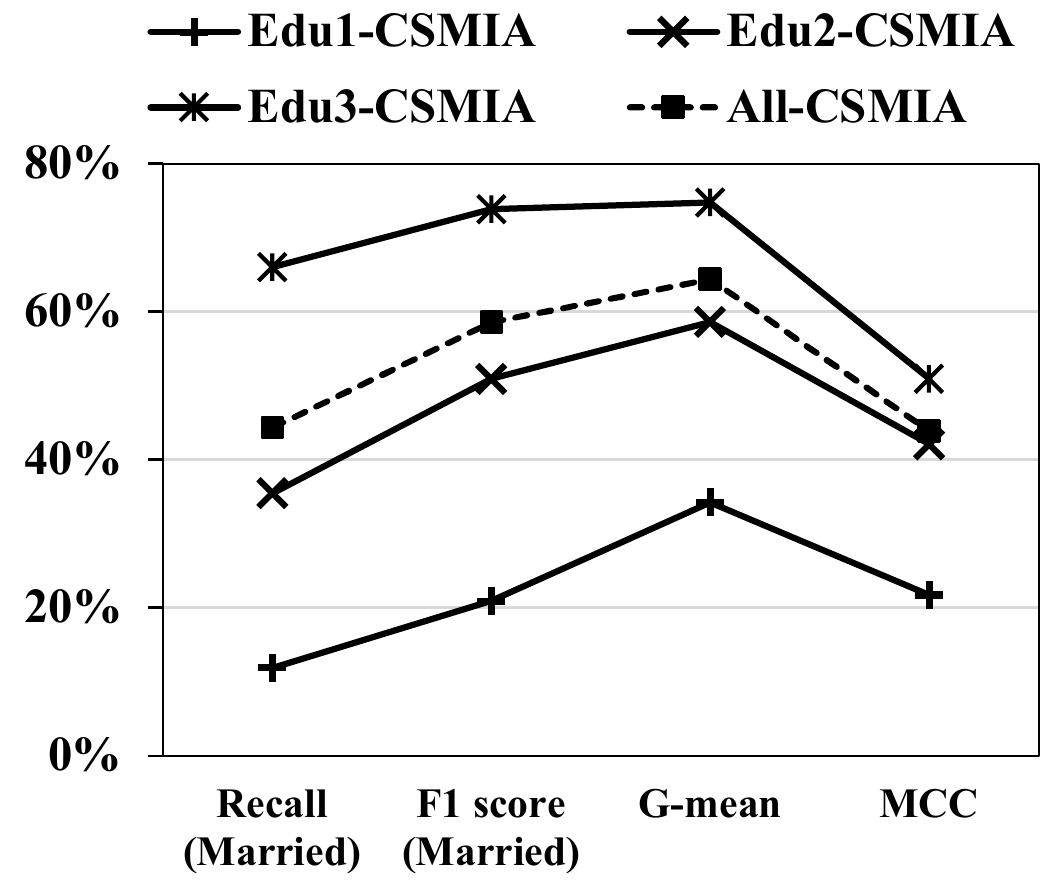}
\label{fig:edu_dnn_csmia}
}
\hfill
\subfigure[CMMIA on decision tree model]{\includegraphics[width=0.48\columnwidth, height=4cm]
{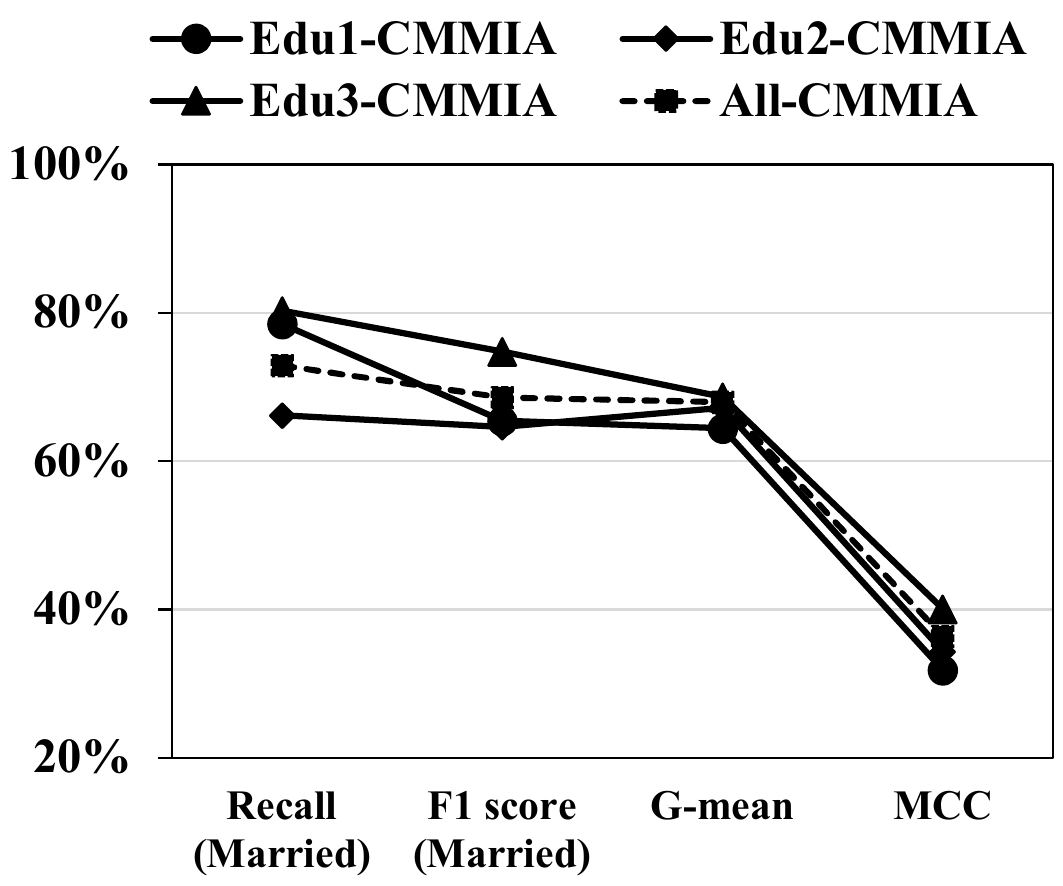}
\label{fig:edu_dt_cmmia}
}
\hfill
\subfigure[CMMIA  on deepnet model]{\includegraphics[width=0.48\columnwidth, height=4cm]
{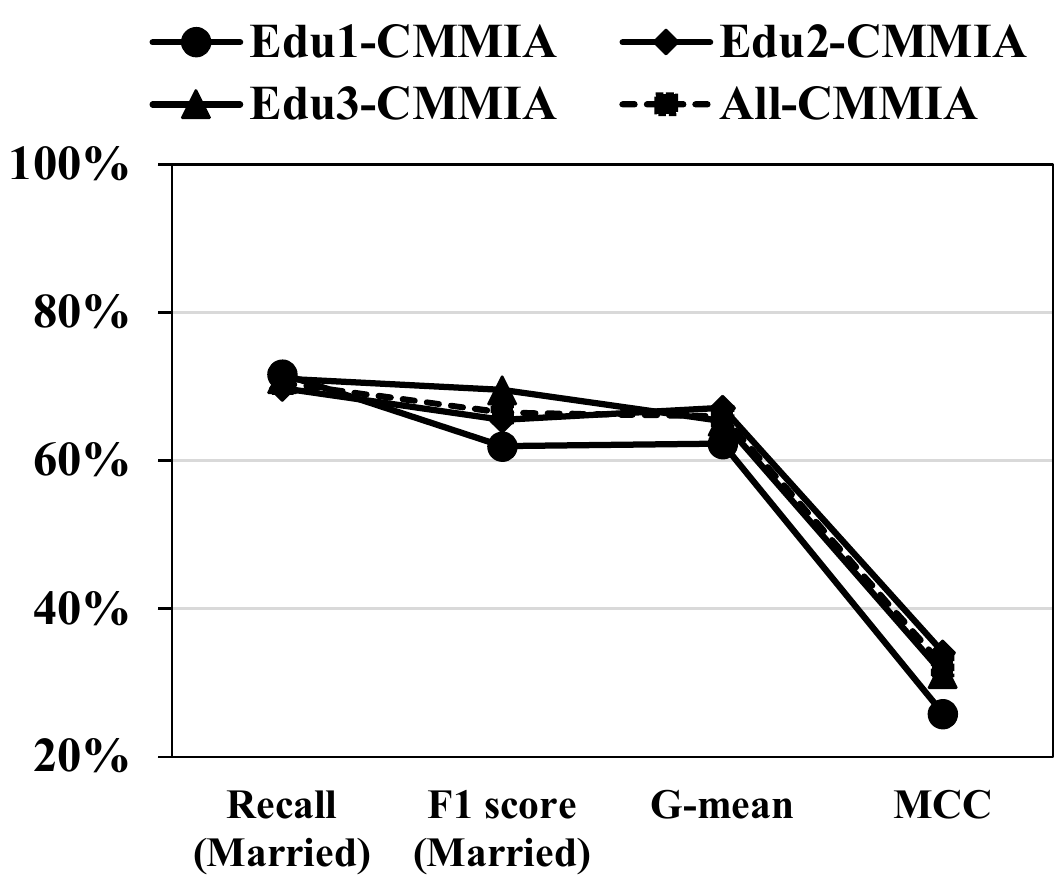}
\label{fig:edu_dnn_cmmia}
}
\caption{Differences in attack performances against individuals with different education levels in the Adult dataset. Edu1: \{Preschool-12th\}, Edu2: \{HS-grad, Some-college\}, Edu3: \{Assoc-voc,  Assoc-acdm, Bachelors, Masters, Prof-school, Doctorate\}.}
\label{fig:edu}
\end{figure*}

\subsection{Differences in Vulnerability Among Different Groups}
\label{subsec:expres_target}
In this section, we further investigate the vulnerability of model inversion attacks by analyzing the attack performances on different groups in the dataset. If a particular group in a dataset is more vulnerable to model inversion attacks than others, it raises serious privacy concerns for that particular group. For instance, we studied the vulnerability of individuals in the Adult dataset by grouping them according to their increasing education levels. We clustered them into three groups, Edu1: \{Preschool-12th\}, Edu2: \{HS-grad, Some-college\}, and Edu3: \{Assoc-voc,  Assoc-acdm, Bachelors, Masters, Prof-school, Doctorate\}. The clustered data is then more balanced with number of instances 4397, 19169, and 11656, respectively. The percentages of married individuals in these groups are $43.39\%$, $45.01\%$, and $54.55\%$, respectively. 
Fig.~\ref{fig:edu} shows the contrast of attack performances against these three groups.

Figures~\ref{fig:edu_dt_csmia} and~\ref{fig:edu_dnn_csmia} show how the CSMIA attack performs on these three groups of individuals where the target models are decision tree and deep neural network, respectively. As demonstrated in the figures, the Edu3 group is the most vulnerable one with the highest recall, F1 score, G-mean, and MCC, independent of the machine learning algorithm used to train the target model. We also demonstrate the performance of CSMIA on the entire dataset (all of Edu1, Edu2, and Edu3 combined) with dotted lines in Figures~\ref{fig:edu_dt_csmia} and~\ref{fig:edu_dnn_csmia}. 

Note that the performance of an adversary with RandGA strategy would not differ significantly for these different groups because of their random prediction. Due to the differences in the underlying distributions of the married individuals in these groups, the RandGA strategy would only show slightly different performance in terms of precision and thus in the F1 score. 
While this result shows only one instance of such inequity in the model inversion attack performances on different groups, this is a potentially serious issue and needs to be further investigated.  Otherwise, while it may seem that the attack performance on the overall dataset is not a significant threat, some specific groups in the dataset could be significantly more vulnerable to model inversion attacks.

The CMMIA strategy's performance on these three groups of individuals does not vary significantly as shown in Figures~\ref{fig:edu_dt_cmmia} and~\ref{fig:edu_dnn_cmmia}. This confirms that the CMMIA strategy is capable of better handling the sensitivity of the target model in terms of the returned confidence scores when the attributes, both sensitive and non-sensitive, are varied (also discussed in Sections~\ref{subsubsec:expres_mia_gss} and~\ref{subsec:model_vs_attack}).

\begin{figure*}[t]
\centering
\includegraphics[width=1\textwidth, height=3.9cm]
{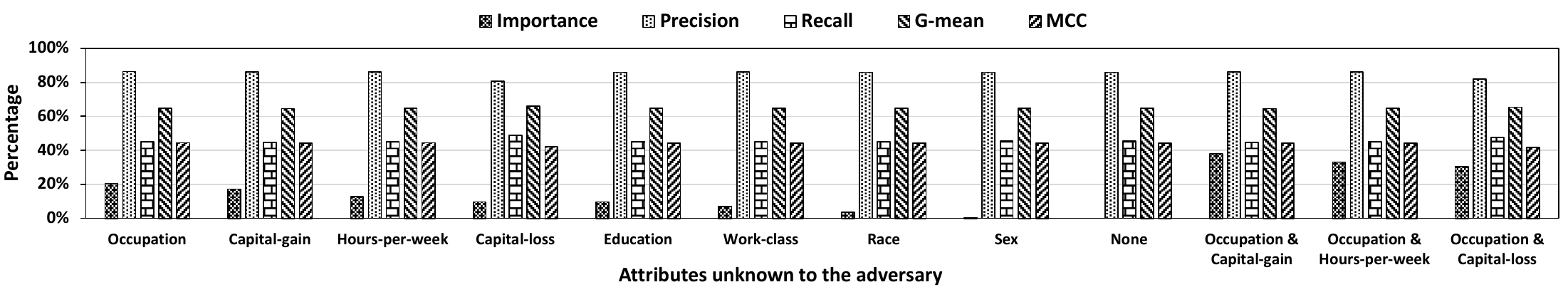}
\caption{CSMIA performance against the decision tree model trained on Adult dataset when some of the other (non-sensitive) attributes of a target individual are also unknown to the adversary.}
\label{fig:missing_adult_dt_cs}
\end{figure*}

\subsection{Model Inversion Attack Results With Partial Knowledge of Target Individual's Non-sensitive Attributes}
\label{subsec:expres_mia_partial}
In Section~\ref{subsec:expres_indepth}, we have observed that the models trained on the Adult dataset are notably more vulnerable to model inversion attacks when compared to the models trained on the GSS dataset. Therefore, in this section, we focus on only the Adult dataset to evaluate how our proposed attacks would perform when the adversary has partial knowledge of the target individual's non-sensitive attributes.

As mentioned earlier, we have removed the `\emph{relationship}' attribute from the Adult dataset due to its very high correlation with the \emph{marital status} sensitive attribute. Excluding the sensitive attribute (`marital status') and the output of the target model (`income'), we consider each of the remaining (non-sensitive) attributes to be unknown to the adversary once at a time, i.e., denoting those as $x_2$. Fig.~\ref{fig:missing_adult_dt_cs} shows the performance of our confidence score-based attack on the decision tree target model trained on the Adult dataset when some of the non-sensitive attributes are unknown to the adversary. The x-axis shows the non-sensitive attributes that are unknown. The attributes are sorted (from left to right) according to their \emph{importance} in the model, a parameter computed by BigML. We also present the original results (i.e., when \emph{none} of the non-sensitive attributes is unknown to the adversary) to compare how the partial knowledge of the target individual's non-sensitive attributes impacts our attacks' performances. As demonstrated in Fig.~\ref{fig:missing_adult_dt_cs}, we observe that the performance of our attack does not deteriorate and remain almost same when some of the non-sensitive attributes are unknown to the adversary, independent of the importance of the attributes in the target model. We observe only slightly lower precision (and slightly higher recall) when the `capital-loss' attribute is unknown to the adversary. We also perform experiments where a combination of non-sensitive attributes are unknown to the adversary-- `occupation and capital-gain' (combined importance $37.8\%$),  `occupation and hours-per-week' (combined importance $33.3\%$), and `occupation and capital-loss' (combined importance $30.4\%$). As demonstrated in Fig.~\ref{fig:missing_adult_dt_cs}, our attack does not show any significant deterioration. \emph{These results not only show an increased vulnerability of  model inversion attacks but also escalate the practicability of such attacks in the real world where the adversary may not know all other attributes of a target individual.}

Due to space constraints and also due to similarities with the results in Fig.~\ref{fig:missing_adult_dt_cs}, the performance details against the deepnet target model in this setting have been discussed in Appendix~\ref{appnB}.

\begin{table*}[t]
  \centering
  \caption{Attack performance against the deep neural network target model trained on Adult dataset.}
  \resizebox{.84\textwidth}{!}{
  \begin{tabular}{ | l | l | l | l | l | l | l | l | l | l |}
    \hline
    Target model & \multirow{2}{*}{Attack Strategy} &  \multirow{2}{*}{TP} & \multirow{2}{*}{TN} & \multirow{2}{*}{FP} & \multirow{2}{*}{FN} & \multirow{2}{*}{Precision} & \multirow{2}{*}{Recall} & \multirow{2}{*}{Accuracy} & \multirow{2}{*}{F1 score}  \\
    class label & & & & & & & & &  \\ \hline 
    
    \multirow{3}{*} {<=50K} & FJR attack~\cite{FredriksonCCS2015} & $13$ & $17108$ & $13$ & $9315$ & $50\%$ & $00.14\%$ & $64.73\%$ & $00.28\%$  \\ 
    \cline{2-10} & Confidence score-based attack & $127$ & $17018$ & $103$ & $9201$ & $55.22\%$ & $1.36\%$ & $64.82\%$ & $2.66\%$ \\ 
    \cline{2-10} & Confidence modeling-based attack & $5643$ & $11399$ & $5722$ & $3685$ & $49.65\%$ & $60.5\%$ & $64.43\%$ & $54.54\%$  \\\hline 
    
    \multirow{3}{*} {>50K} & FJR attack~\cite{FredriksonCCS2015} & $3775$ & $710$ & $498$ & $3790$ & $88.34\%$ & $49.9\%$ & $51.12\%$ & $63.78\%$  \\
    \cline{2-10} & Confidence score-based attack & $7537$ & $67$ & $1141$ & $28$ & $86.85\%$ & $99.63\%$ & $86.68\%$ & $92.8\%$ \\ 
    \cline{2-10} & Confidence modeling-based attack & $6668$ & $220$ & $988$ & $897$ & $88\%$ & $88.14\%$ & $78.51\%$ & $87.62\%$  \\ \hline

  \end{tabular}
  }
\label{table:class_label}
\end{table*}

\begin{table}[h]
  \centering
  \caption{Confusion matrix of decision tree target model trained on Adult dataset.}
  \resizebox{0.49\textwidth}{!}{
\begin{tabular}{|l|*{4}{c|}}\hline
    \backslashbox{Actual}{Predicted}
    & <=50K & >50K  & Total & Recall \\\hline
    <=50K & 24912 & 1537 & 26449 & 94.19\% \\\hline
    >50K  & 3343 & 5430 & 8773 & 61.89\%  \\\hline
    Total & 28255 & 6967 & 35222  & Avg. recall 78.04\% \\ \hline
    Precision  & 88.17\% & 77.94\%  & Avg. precision 83.05\% & Accuracy 86.15\% \\ \hline
\end{tabular}
  }
\label{table:CM_DT_Adult_main}
\end{table}

\subsection{Model Inversion Attacks' Efficacy on Different Class Labels of Target Model}
\label{subsec:model_vs_attack}
In this section, we aim to understand the efficacy of model inversion attacks for different class labels of the target model  and focus on the decision tree target model trained on Adult dataset (Table~\ref{table:class_label}).

Table~\ref{table:CM_DT_Adult_main} shows the confusion matrix of this target model. From the confusion matrix, it is evident that the model's performance is better for class label $<=50K$ than that of class label $>50K$. 

Table~\ref{table:class_label} shows a comparison among FJR attack~\cite{FredriksonCCS2015}, CMMIA, and CSMIA performances for different class labels of the target model. For both FJR attack~\cite{FredriksonCCS2015} and CSMIA, the performance of the attacks are significantly different for the two class labels, e.g., the recall of identifying `married' individuals in class <=50K is very low compared to the recall of identifying `married' individuals in class >50K. In contrast, the performance of CMMIA strategy in identifying `married' individuals is more consistent across the two class labels.

The reason behind the low recall of FJR attack~\cite{FredriksonCCS2015} for class <=50K is that when the adversary queries the model with records from class <=50K by varying sensitive attribute values, i.e., single and married, most of the time both the queries predict the correct class <=50K. In other words, both the query outcomes fall into the [<=50K, <=50K] cell of the confusion matrix. 
Therefore, when the FJR attack~\cite{FredriksonCCS2015} computes $\mathcal{C}[y, y'] * p_{1,i}$ for each possible sensitive attribute value, the $\mathcal{C}[y, y']$ term for both the sensitive attribute values become same. This attack then relies on the $p_{1,i}$ term for attack prediction and thus mostly predicts `single' because of its higher marginal prior  which \emph{simply boils the FJR attack~\cite{FredriksonCCS2015} down to a naive attack}. However, for the records with class label >50K, the FJR attack~\cite{FredriksonCCS2015} performs better since queries with varying sensitive attribute values produce target model outcomes that fall into different cells of the confusion matrix. The $\mathcal{C}[y, y']$ term dominates $p_{1,i}$ and thus the attack returns more true positives. 

The CSMIA strategy also shows similar trends based on the confidence scores returned by the target model, i.e., the performance of the attack is significantly different for the two class labels. However, CSMIA outperforms FJR attack~\cite{FredriksonCCS2015} significantly in terms of recall for class >50K.

Unlike FJR attack~\cite{FredriksonCCS2015} and CSMIA, the CMMIA strategy with trained attack models performs well on <=50K class instances with a recall of $60.5\%$ while achieving a precision similar to the other attacks. The CMMIA strategy also performs well on >50K class instances with significantly high recall when compared to FJR attack~\cite{FredriksonCCS2015}. These results show the advantage of the CMMIA strategy over the other attacks which is clearly due to the fact that the CMMIA strategy focuses on learning the correlation between the outcome (both predictions and confidence scores) of the target model and the sensitive attribute by training the attack models.

\subsection{Discussion}
\label{subsec:discussion}
As mentioned in Section~\ref{sec:intro}, the TIR attacks have strong correlations with the model's predictive power. This is because highly predictive models are able to establish a strong correlation between features and labels, and this is the property that an adversary exploits to mount the TIR attacks~\cite{Zhang_2020_CVPR}. However, we argue that such is not the case for MIAI attacks. In Section~\ref{subsec:model_vs_attack}, where it is evident that the target model's performance is better for class label $<=50K$ than that of class label $>50K$, we demonstrate that the MIAI attacks perform better against the records of class label $>50K$. Moreover, in Section~\ref{subsec:expres_mia}, we show that the correlations between the sensitive attributes and the corresponding target models trained on GSS and Adult datasets do not differ significantly whereas the proposed MIAI attacks on these target models demonstrate significantly different results (attacks against target models trained on Adult dataset are more effective than that of against the target models trained on GSS dataset). This indicates that only controlling the \emph{importance} of the sensitive attributes in the target model may not be always sufficient to reduce the risk of model inversion attacks. Finally, we investigate if black-box access to a particular model really helps the adversary to estimate the sensitive attributes in the training dataset which is otherwise impossible for the adversary to estimate, i.e., without access to the black-box model, by performing extensive experiment with baseline attacks (random guessing attack and naive attack). 

Hence, \emph{ours is the first work that studies the MIAI attacks in such details} and it is evident that further investigation is required to better understand the potentially serious threats of model inversion attacks such as its unequal impact on different groups of individuals as described in Section~\ref{subsec:expres_target}.

\section{Related Work}
\label{sec:related}

In~\cite{PharmaUSENIX2014}, Fredrikson et al. introduced the concept of model inversion attacks and applied their attack to linear regression models. In~\cite{FredriksonCCS2015}, Fredrikson et al. extended their attack so that it could also be applicable to non-linear models, such as decision trees. The later work presents two types of applications of the model inversion attack. The first one assumes an adversary who has access to a model (for querying) and aims to learn the sensitive attributes in the dataset that has been used to train that model (also known as attribute inference attack). In the second setting, the adversary aims to reconstruct instances similar to ones in the training dataset using gradient descent. Particularly, their attack generates images similar to faces used to train a facial recognition model. As mentioned earlier, we focus on the first one, i.e., attribute inference attack. Subsequently, Wu et al.~\cite{wu2016methodology} presented a methodology to formalize the model inversion attack.

A number of attribute inference attacks have been shown to be effective in different domains, such as social media~\cite{GongTOPS2018, AttriInferWWW2017, GongUSENIX2016, GongTIST2014, KosinskiPNAS2013, ChaabaneNDSS2012, SNPrivacyWWW2009} and recommender systems~\cite{weinsberg2012blurme, wu2020joint}. In the case of social media, the adversary infers the private attributes of a user (e.g., gender, political views, locations visited) by leveraging the knowledge of other attributes of that same user that are shared publicly (e.g., list of pages liked by the user, etc). The adversary first trains a machine learning classifier that takes as input the public attributes and then outputs the private attributes. In order to build such a classifier, these attacks~\cite{GongTOPS2018, AttriInferWWW2017, GongUSENIX2016, GongTIST2014, KosinskiPNAS2013, ChaabaneNDSS2012, SNPrivacyWWW2009} rely on social media users who also make their private attributes public. 
Also, for the attacks shown in the recommender systems~\cite{weinsberg2012blurme}, at first, the adversary  collects data of the users who also share their private attributes (e.g., gender) publicly along with their public rating scores (e.g., movie ratings). While our CMMIA strategy assumes the adversary to be able to obtain a dataset from the same population the target model training dataset has been obtained from, in our CSMIA attack we do not make such an assumption. This is because in some special scenarios such an assumption (adversary having access to a similar dataset) may not be valid and our goal is to also incorporate these scenarios in our attack surface so that our attack could be applied more widely.

Shokri et al.~\cite{shokri2019privacy} investigate whether transparency of machine learning models conflicts with privacy and demonstrate that record-based explanations of machine learning models can be effectively exploited by an adversary to reconstruct the training dataset. In their setting, the adversary can generate unlimited transparency queries and for each query, the adversary is assumed to get in return some of the original training dataset records (that are related to the queries) as part of the transparency report. 
He et al.~\cite{HeACSAC2019} devise a new set of model inversion attacks against collaborative inference where a deep neural network and the corresponding inference task are distributed among different participants. The adversary, as a malicious participant, can accurately recover an arbitrary input fed into the model, even if it has no access to other participants' data or computations, or to prediction APIs to query the model. 

Most of the work mentioned above assume that the attributes of a target individual, except the sensitive attribute, are known to the adversary. Hidano et al.~\cite{HidanoPST2017} proposed a method to infer the sensitive attributes without the knowledge of non-sensitive attributes. However, they consider an online machine learning model and assume that the adversary has the capability to poison the model with malicious training data. In contrast, our model inversion attack with partial knowledge of target individual's non-sensitive attributes does not require poisoning and performs similar to scenarios where the adversary has full knowledge of target individual's non-sensitive attributes.

Zhang et al.~\cite{Zhang_2020_CVPR} present a generative model-inversion attack to invert deep neural networks. They demonstrate the effectiveness of their attack by reconstructing face images from a state-of-the-art face recognition classifier. They also prove that a model’s predictive power and its vulnerability to inversion attacks are closely related, i.e., highly predictive models are more vulnerable to inversion attacks.
Aïvodji et al.~\cite{avodji2019gamin} introduce a new black-box model inversion attack framework,
GAMIN (Generative Adversarial Model INversion), based on the continuous training of
a surrogate model for the target model and  evaluate their attacks on
convolutional neural networks. In~\cite{YangCCS2019}, Yang et al. train a second neural network that acts as the inverse of the target model while assuming partial knowledge about the target model's training data. The objective of the works mentioned above is typical instance reconstruction (TIR), i.e., similar to the second attack mentioned in~\cite{FredriksonCCS2015}.

\section{Conclusion and Future Work}
In this paper, we demonstrate two new black-box model inversion attribute inference (MIAI) attacks: (1) confidence modeling-based attack (CMMIA) and (2) confidence score-based attack (CSMIA). The CMMIA strategy assumes the adversary has access to a dataset from the same population the target model training dataset has been obtained from (similar to other state-of-the-art attacks~\cite{YangCCS2019, GongTOPS2018, AttriInferWWW2017, GongUSENIX2016, GongTIST2014, KosinskiPNAS2013, ChaabaneNDSS2012, SNPrivacyWWW2009, weinsberg2012blurme}) whereas the CSMIA strategy does not make such an assumption. Along with accuracy and F1 score, we propose to use the G-mean and Matthews correlation coefficient (MCC) metrics in order to ensure effective evaluation of our attacks as well as the state-of-the-art attacks. We perform an extensive evaluation of our attacks using two types of machine learning models, decision tree and deep neural network, that are trained with two real datasets~\cite{gss, adult}.  Our evaluation results show that the proposed attacks significantly outperform the existing ones. Moreover, we empirically show that model inversion attacks have inequity property and consequently, a particular subset of the training dataset (grouped by attributes, such as gender, education level, etc.) could be more vulnerable than others to the model inversion attacks. We also evaluate the risks incurred by model inversion attacks when the adversary does not have knowledge of all other non-sensitive attributes of the target individual and demonstrate that our attack's performance is not impacted significantly in those scenarios.

Further investigating the potentially serious threats of model inversion attacks such as its inequity property  on different groups of individuals and also, extending the model inversion attacks to dynamic environments where the model and/or the non-sensitive attributes of the target individuals could change over time are left as future work. Since the defense methods designed to mitigate reconstruction of instances resembling those used in the training dataset (TIR attacks)~\cite{alves2019,yang2020defending} do not directly apply to our MIAI attack setting, it would also be an interesting direction for future work to explore new defense methods.

\bibliographystyle{unsrt}
\bibliography{main}

\begin{thebibliography}{10}

\bibitem{medicine1}
International Warfarin~Pharmacogenetics Consortium.
\newblock Estimation of the warfarin dose with clinical and pharmacogenetic
  data.
\newblock {\em New England Journal of Medicine}, 360(8):753--764, 2009.

\bibitem{medicine2}
Jeremy~C Weiss, Sriraam Natarajan, Peggy~L Peissig, Catherine~A McCarty, and
  David Page.
\newblock Machine learning for personalized medicine: Predicting primary
  myocardial infarction from electronic health records.
\newblock {\em Ai Magazine}, 33(4):33--33, 2012.

\bibitem{medicine3}
Davide Cirillo and Alfonso Valencia.
\newblock Big data analytics for personalized medicine.
\newblock {\em Current opinion in biotechnology}, 58:161--167, 2019.

\bibitem{medicine4}
Marinka Zitnik, Francis Nguyen, Bo~Wang, Jure Leskovec, Anna Goldenberg, and
  Michael~M Hoffman.
\newblock Machine learning for integrating data in biology and medicine:
  Principles, practice, and opportunities.
\newblock {\em Information Fusion}, 50:71--91, 2019.

\bibitem{recom1}
Xiaoyuan Su and Taghi~M Khoshgoftaar.
\newblock A survey of collaborative filtering techniques.
\newblock {\em Advances in artificial intelligence}, 2009, 2009.

\bibitem{recom2}
G.~{Linden}, B.~{Smith}, and J.~{York}.
\newblock Amazon.com recommendations: item-to-item collaborative filtering.
\newblock {\em IEEE Internet Computing}, 7(1):76--80, 2003.

\bibitem{recom3}
Yehuda Koren, Robert Bell, and Chris Volinsky.
\newblock Matrix factorization techniques for recommender systems.
\newblock {\em Computer}, 42(8):30--37, 2009.

\bibitem{finlaw1}
Christian Dunis, Peter~W Middleton, A~Karathanasopolous, and K~Theofilatos.
\newblock {\em Artificial intelligence in financial markets}.
\newblock Springer, 2016.

\bibitem{finlaw2}
Robert~R Trippi and Efraim Turban.
\newblock {\em Neural networks in finance and investing: Using artificial
  intelligence to improve real world performance}.
\newblock McGraw-Hill, Inc., 1992.

\bibitem{finlaw3}
Mireille Hildebrandt.
\newblock Law as computation in the era of artificial legal intelligence:
  Speaking law to the power of statistics.
\newblock {\em University of Toronto Law Journal}, 68(supplement 1):12--35,
  2018.

\bibitem{social1}
Daniel Gayo-Avello, Panagiotis~Takis Metaxas, Eni Mustafaraj, Markus
  Strohmaier, Harald Schoen, and Peter Gloor.
\newblock The power of prediction with social media.
\newblock {\em Internet Research}, 2013.

\bibitem{social2}
Golnoosh Farnadi, Geetha Sitaraman, Shanu Sushmita, Fabio Celli, Michal
  Kosinski, David Stillwell, Sergio Davalos, Marie-Francine Moens, and Martine
  De~Cock.
\newblock Computational personality recognition in social media.
\newblock {\em User modeling and user-adapted interaction}, 26(2-3):109--142,
  2016.

\bibitem{social3}
Marcin Skowron, Marko Tkal{\v{c}}i{\v{c}}, Bruce Ferwerda, and Markus Schedl.
\newblock Fusing social media cues: personality prediction from twitter and
  instagram.
\newblock In {\em Proceedings of the 25th international conference companion on
  world wide web}, pages 107--108, 2016.

\bibitem{FredriksonCCS2015}
Matt Fredrikson, Somesh Jha, and Thomas Ristenpart.
\newblock Model inversion attacks that exploit confidence information and basic
  countermeasures.
\newblock In {\em Proceedings of the 22Nd ACM SIGSAC Conference on Computer and
  Communications Security}, CCS '15, pages 1322--1333, New York, NY, USA, 2015.
  ACM.

\bibitem{PharmaUSENIX2014}
Matthew Fredrikson, Eric Lantz, Somesh Jha, Simon Lin, David Page, and Thomas
  Ristenpart.
\newblock Privacy in pharmacogenetics: An end-to-end case study of personalized
  warfarin dosing.
\newblock In {\em 23rd {USENIX} Security Symposium ({USENIX} Security 14)},
  pages 17--32, San Diego, CA, August 2014. {USENIX} Association.

\bibitem{avodji2019gamin}
Ulrich Aïvodji, Sébastien Gambs, and Timon Ther.
\newblock Gamin: An adversarial approach to black-box model inversion.
\newblock 2019.

\bibitem{YangCCS2019}
Ziqi Yang, Jiyi Zhang, Ee-Chien Chang, and Zhenkai Liang.
\newblock Neural network inversion in adversarial setting via background
  knowledge alignment.
\newblock In {\em Proceedings of the 2019 ACM SIGSAC Conference on Computer and
  Communications Security}, CCS ’19, page 225–240, New York, NY, USA, 2019.
  Association for Computing Machinery.

\bibitem{Zhang_2020_CVPR}
Yuheng Zhang, Ruoxi Jia, Hengzhi Pei, Wenxiao Wang, Bo~Li, and Dawn Song.
\newblock The secret revealer: Generative model-inversion attacks against deep
  neural networks.
\newblock In {\em Proceedings of the IEEE/CVF Conference on Computer Vision and
  Pattern Recognition (CVPR)}, June 2020.

\bibitem{gss}
The general social survey.
\newblock https://gss.norc.org/.

\bibitem{adult}
Adult dataset.
\newblock http://archive.ics.uci.edu/ml/datasets/Adult.

\bibitem{Sun2009ClassificationOI}
Yanmin Sun, Andrew K.~C. Wong, and Mohamed~S. Kamel.
\newblock Classification of imbalanced data: a review.
\newblock {\em Int. J. Pattern Recognit. Artif. Intell.}, 23:687--719, 2009.

\bibitem{MATTHEWS1975442}
B.W. Matthews.
\newblock Comparison of the predicted and observed secondary structure of t4
  phage lysozyme.
\newblock {\em Biochimica et Biophysica Acta (BBA) - Protein Structure},
  405(2):442 -- 451, 1975.

\bibitem{bigml}
Bigml.
\newblock https://bigml.com/.

\bibitem{wu2016methodology}
Xi~Wu, Matthew Fredrikson, Somesh Jha, and Jeffrey~F Naughton.
\newblock A methodology for formalizing model-inversion attacks.
\newblock In {\em 2016 IEEE 29th Computer Security Foundations Symposium
  (CSF)}, pages 355--370. IEEE, 2016.

\bibitem{GongTOPS2018}
Neil~Zhenqiang Gong and Bin Liu.
\newblock Attribute inference attacks in online social networks.
\newblock {\em ACM Trans. Priv. Secur.}, 21(1), January 2018.

\bibitem{AttriInferWWW2017}
Jinyuan Jia, Binghui Wang, Le~Zhang, and Neil~Zhenqiang Gong.
\newblock Attriinfer: Inferring user attributes in online social networks using
  markov random fields.
\newblock In {\em Proceedings of the 26th International Conference on World
  Wide Web}, WWW ’17, page 1561–1569, Republic and Canton of Geneva, CHE,
  2017. International World Wide Web Conferences Steering Committee.

\bibitem{GongUSENIX2016}
Neil~Zhenqiang Gong and Bin Liu.
\newblock You are who you know and how you behave: Attribute inference attacks
  via users{\textquoteright} social friends and behaviors.
\newblock In {\em 25th {USENIX} Security Symposium ({USENIX} Security 16)},
  pages 979--995, Austin, TX, August 2016. {USENIX} Association.

\bibitem{GongTIST2014}
Neil~Zhenqiang Gong, Ameet Talwalkar, Lester Mackey, Ling Huang, Eui
  Chul~Richard Shin, Emil Stefanov, Elaine~(Runting) Shi, and Dawn Song.
\newblock Joint link prediction and attribute inference using a
  social-attribute network.
\newblock {\em ACM Trans. Intell. Syst. Technol.}, 5(2), April 2014.

\bibitem{KosinskiPNAS2013}
Michal Kosinski, David Stillwell, and Thore Graepel.
\newblock Private traits and attributes are predictable from digital records of
  human behavior.
\newblock {\em Proceedings of the National Academy of Sciences},
  110(15):5802--5805, 2013.

\bibitem{ChaabaneNDSS2012}
Abdelberi Chaabane, Gergely Acs, and Mohamed~Ali Kaafar.
\newblock You are what you like! information leakage through users’
  interests.
\newblock In {\em In NDSS}, 2012.

\bibitem{SNPrivacyWWW2009}
Elena Zheleva and Lise Getoor.
\newblock To join or not to join: The illusion of privacy in social networks
  with mixed public and private user profiles.
\newblock In {\em Proceedings of the 18th International Conference on World
  Wide Web}, WWW ’09, page 531–540, New York, NY, USA, 2009. Association
  for Computing Machinery.

\bibitem{weinsberg2012blurme}
Udi Weinsberg, Smriti Bhagat, Stratis Ioannidis, and Nina Taft.
\newblock Blurme: Inferring and obfuscating user gender based on ratings.
\newblock In {\em Proceedings of the sixth ACM conference on Recommender
  systems}, pages 195--202, 2012.

\bibitem{wu2020joint}
Le~Wu, Yonghui Yang, Kun Zhang, Richang Hong, Yanjie Fu, and Meng Wang.
\newblock Joint item recommendation and attribute inference: An adaptive graph
  convolutional network approach.
\newblock 2020.

\bibitem{shokri2019privacy}
Reza Shokri, Martin Strobel, and Yair Zick.
\newblock Privacy risks of explaining machine learning models.
\newblock {\em arXiv preprint arXiv:1907.00164}, 2019.

\bibitem{HeACSAC2019}
Zecheng He, Tianwei Zhang, and Ruby~B. Lee.
\newblock Model inversion attacks against collaborative inference.
\newblock In {\em Proceedings of the 35th Annual Computer Security Applications
  Conference}, ACSAC ’19, page 148–162, New York, NY, USA, 2019.
  Association for Computing Machinery.

\bibitem{HidanoPST2017}
S.~{Hidano}, T.~{Murakami}, S.~{Katsumata}, S.~{Kiyomoto}, and G.~{Hanaoka}.
\newblock Model inversion attacks for prediction systems: Without knowledge of
  non-sensitive attributes.
\newblock In {\em 2017 15th Annual Conference on Privacy, Security and Trust
  (PST)}, pages 115--11509, 2017.

\bibitem{alves2019}
Tiago A.~O. Alves, Felipe M.~G. Fran\c{c}a, and Sandip Kundu.
\newblock Mlprivacyguard: Defeating confidence information based model
  inversion attacks on machine learning systems.
\newblock In {\em Proceedings of the 2019 on Great Lakes Symposium on VLSI},
  GLSVLSI ’19, page 411–415, New York, NY, USA, 2019. Association for
  Computing Machinery.

\bibitem{yang2020defending}
Ziqi Yang, Bin Shao, Bohan Xuan, Ee-Chien Chang, and Fan Zhang.
\newblock Defending model inversion and membership inference attacks via
  prediction purification, 2020.

\end{thebibliography}

\appendix
\section{Appendix}

\begin{figure*}[t]
\centering
\subfigure[Marginal prior of the positive class attribute is $0.3$.]{\includegraphics[width=0.32\textwidth, height=5cm]
{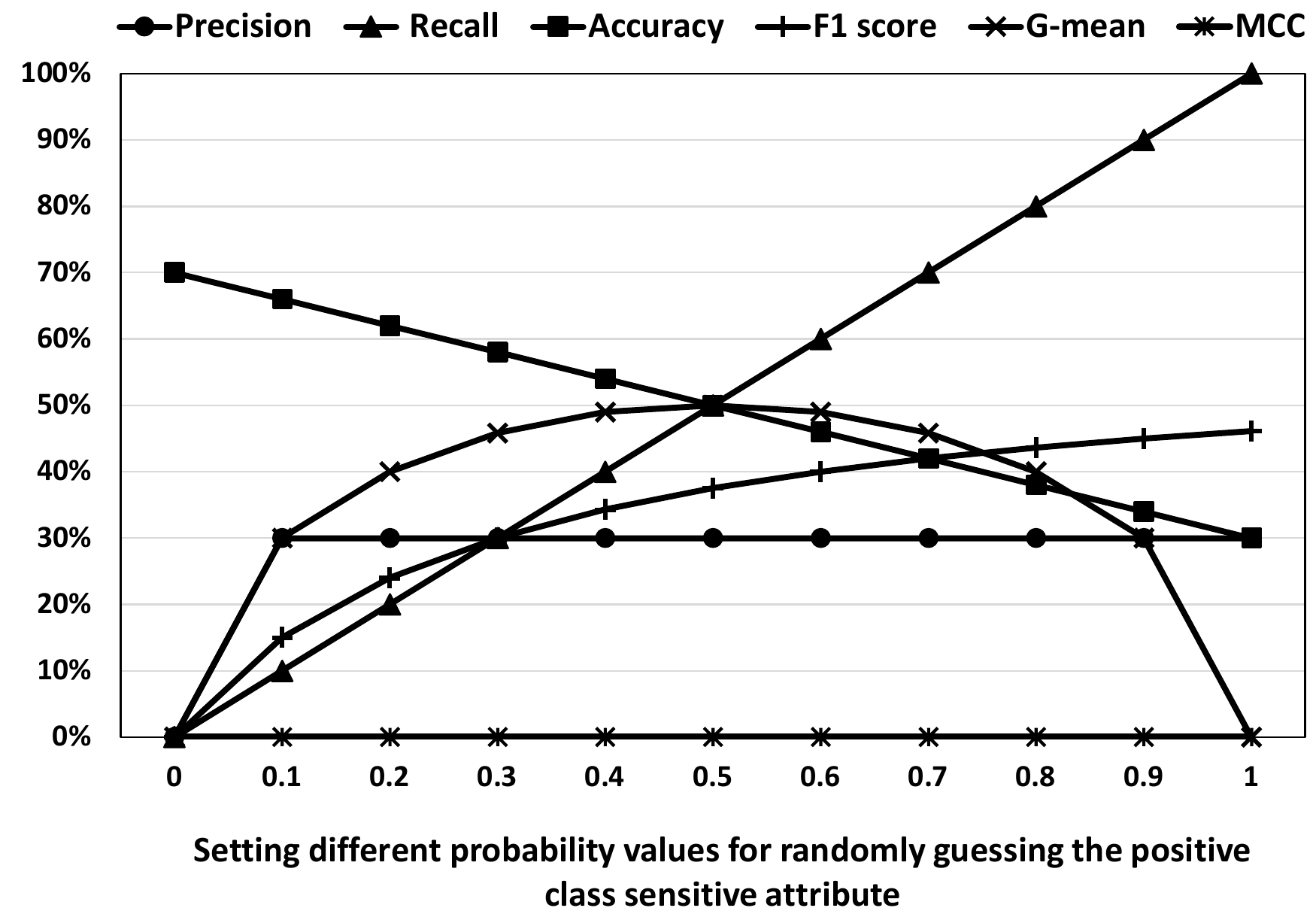}
\label{fig:random_30}
}
\hfill
\subfigure[GSS dataset where the marginal prior of the positive class attribute is $0.197$.]{\includegraphics[width=0.32\textwidth, height=5cm]
{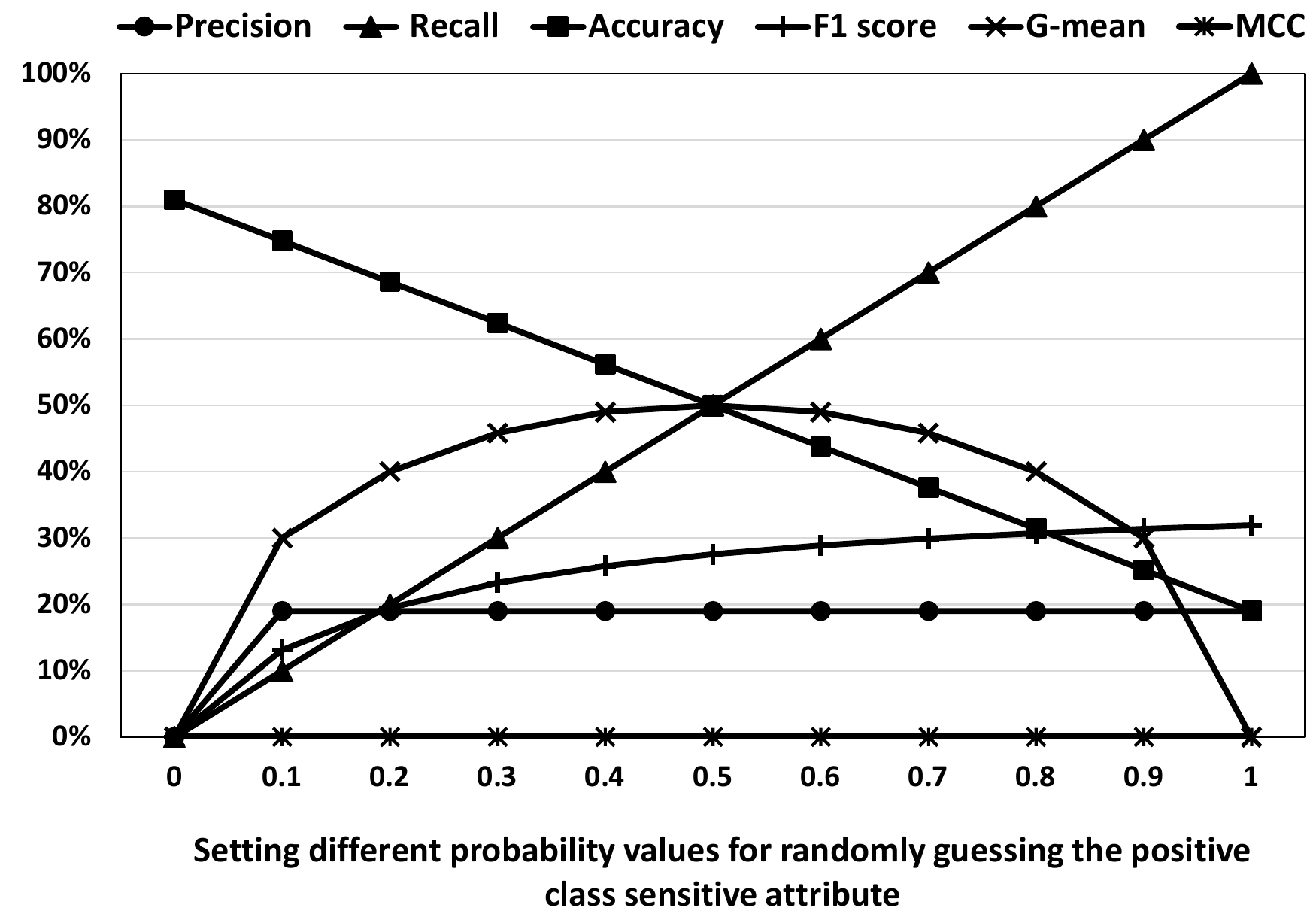}
\label{fig:random_gss}
}
\hfill
\subfigure[Adult dataset where the marginal prior of the positive class attribute is $0.479$.]{\includegraphics[width=0.32\textwidth, height=5cm]
{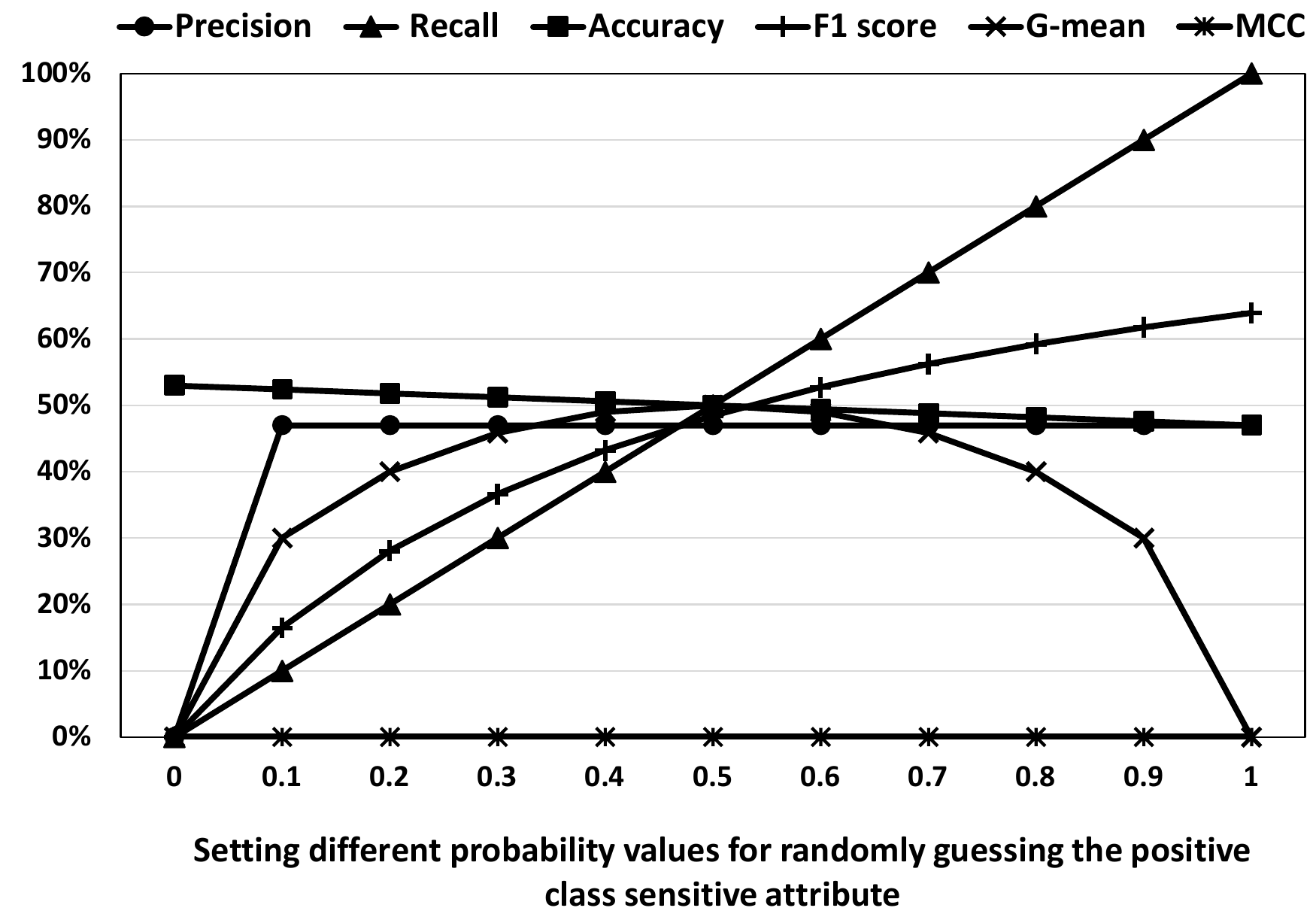}
\label{fig:random_adult}
}
\caption{Random guessing attack performances for different marginal priors of the positive class sensitive attribute value.}
\label{fig:random}
\end{figure*}

\begin{figure*}[h]
\centering
\includegraphics[width=0.97\textwidth, height=3.5cm]
{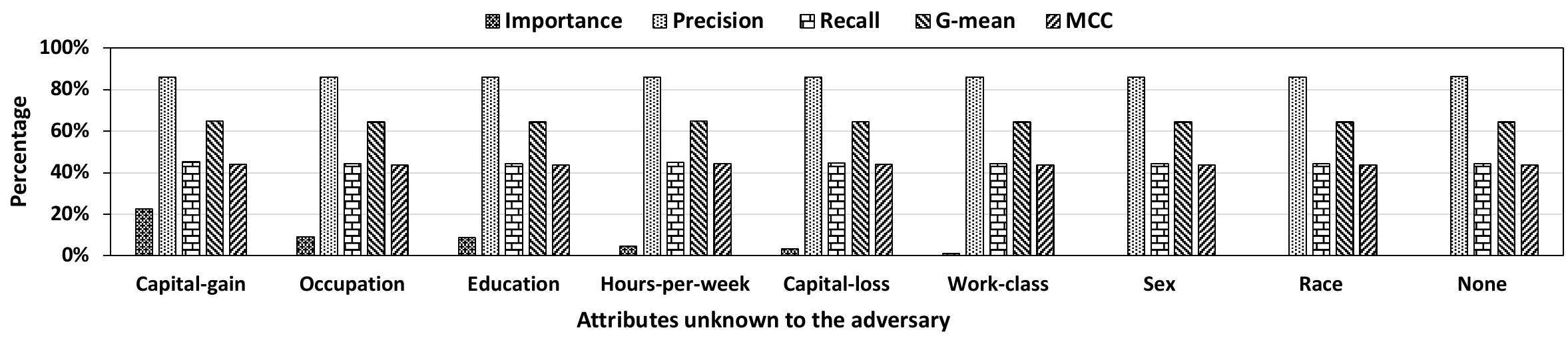}
\caption{Confidence score-based attack performance against the deep neural network model trained on Adult dataset when some of the other (non-sensitive) attributes of a target individual are also unknown to the adversary.}
\label{fig:missing_adult_dnn_cs}
\end{figure*}

\subsection{Random Guessing Attack Performances}
\label{appn:random}
In this attack, the adversary randomly predicts the sensitive attribute by setting a probability for the positive class sensitive attribute value. Fig.~\ref{fig:random_30} shows the optimal performance of random guessing attack when the marginal prior of the positive class sensitive attribute is $0.3$ and the adversary sets different probabilities to predict the positive class sensitive attribute value (probabilities in x-axis). As shown in the figure, the maximum G-mean a random guessing attack can achieve is 50\%, independent of the knowledge of marginal prior. The precision for predicting the positive class sensitive attribute is constant and equals the marginal prior of that class as long as the  set probability is $>0$. This is because when the attack randomly assigns positive class label to the records, approximately 30\% of those records' sensitive attributes would turn out to be originally positive according to the marginal prior of the positive class sensitive attribute which is $0.3$. The recall of random guessing attack increases with the probability set to predict the positive class sensitive attribute. For example, if the adversary reports all the records' sensitive attributes as positive, there is no false negative left and thus recall reaches 100\%. Figures~\ref{fig:random_gss} and~\ref{fig:random_adult} show the performance of random guessing attack on the GSS and Adult datasets, respectively, when the adversary sets different probability values to predict the positive class sensitive attribute.
As shown in Figures~\ref{fig:random_30},~\ref{fig:random_gss}, and~\ref{fig:random_adult}, the MCC of the random guessing attacks is always 0.

\subsection{Model Inversion Attack Results With Partial Knowledge of Target Individual's Non-sensitive Attributes}
\label{appnB}
Fig.~\ref{fig:missing_adult_dnn_cs} shows the performance of our confidence score-based attack on the deep neural network target model trained on the Adult dataset when some of the non-sensitive attributes are unknown to the adversary. The x-axis shows the non-sensitive attribute that is unknown. The attributes are sorted (from left to right) according to their \emph{importance} in the model. We also present the original results (i.e., when \emph{none} of the non-sensitive attributes is unknown to the adversary) to compare how the partial knowledge of the target individual's non-sensitive attributes impacts our attacks' performances. 
As demonstrated in the figure, we observe that the performances of our attack do not deteriorate and remain almost the same when some of the non-sensitive attributes are unknown to the adversary, independent of the importance of the attributes in the target model.

\begin{figure*}[t]
\centering
\subfigure[GSS dataset attributes' importance]{\includegraphics[width=0.45\textwidth, height=4cm]
{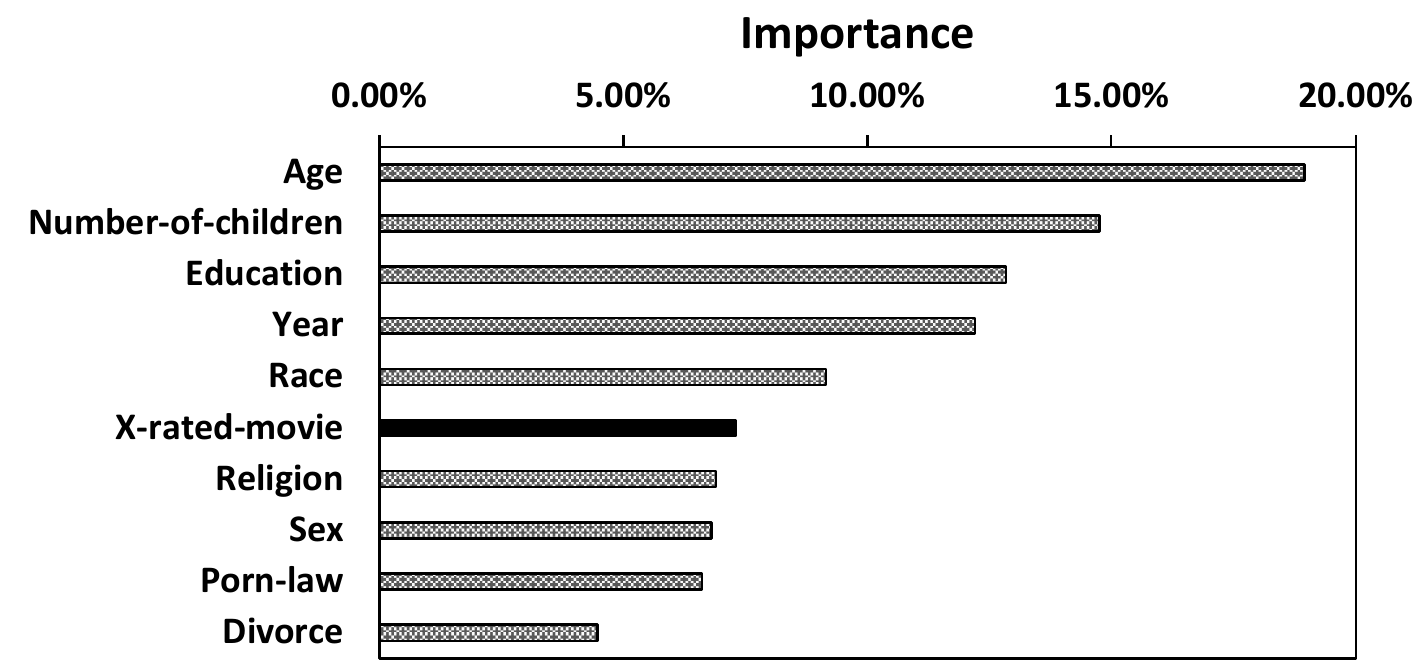}
}
\hfill
\subfigure[Adult dataset attributes' importance]{\includegraphics[width=0.45\textwidth, height=4cm]
{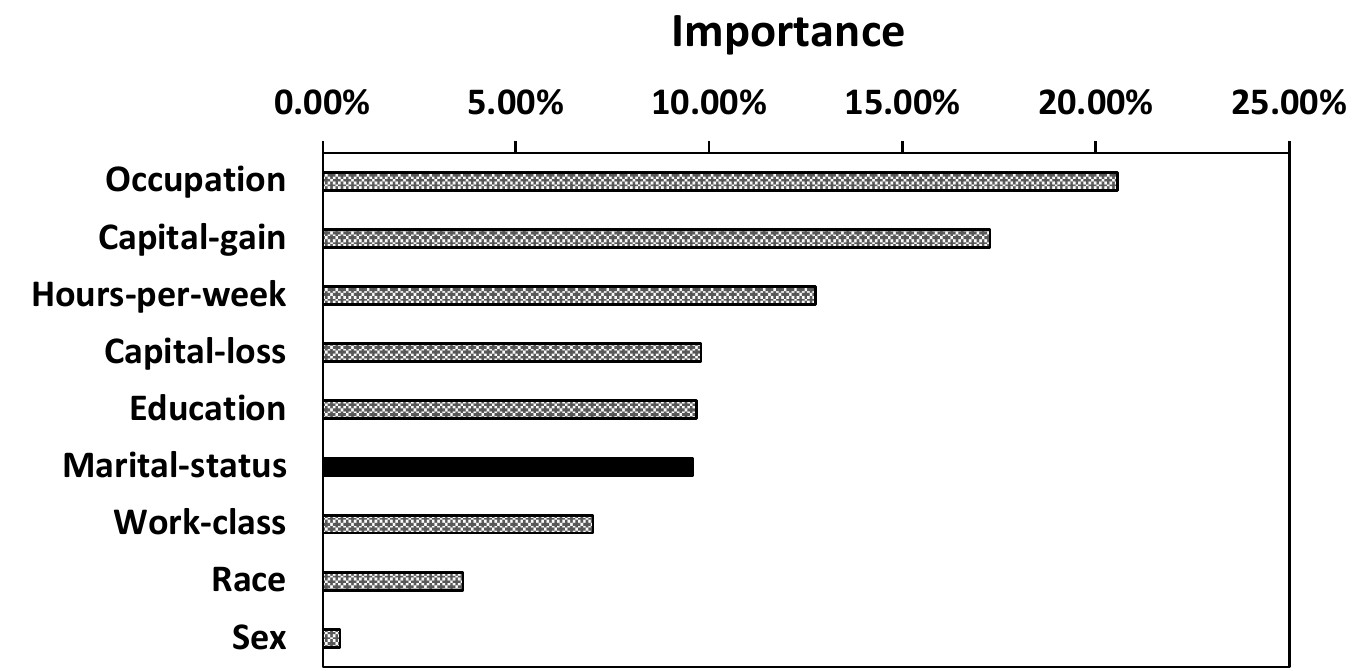}
}
\caption{Importance of GSS and Adult dataset attributes on their corresponding decision tree target models.}
\label{fig:importance}
\end{figure*}

\begin{table*}[h]
  \centering
  \caption{Confusion matrix of decision tree target model trained on GSS dataset.}
  \resizebox{0.75\textwidth}{!}{
\begin{tabular}{|l|*{5}{c|}}\hline
    \backslashbox{Actual}{Predicted}
    &Not too happy& Pretty happy& Very happy & Total & Recall \\\hline
    Not too happy & 5 & 63 & 370 & 438 & 1.14\% \\\hline
    Pretty happy  & 0 & 813 & 4178 & 4991 & 16.29\% \\\hline
    Very happy    & 0 & 526 & 9280 & 9806 & 94.64\% \\\hline
    Total         & 5 & 1402 & 13828 & 15235 & Avg. recall 37.36\% \\ \hline
    Precision     & 100\% & 57.99\% & 67.11\% & Avg. precision 75.03\% & Accuracy 66.28\% \\ \hline
\end{tabular}
  }
\label{table:CM_DT_GSS}
\vspace{0.4cm}

  \centering
  \caption{Confusion matrix of deepnet target model trained on GSS dataset.}
  \resizebox{0.75\textwidth}{!}{
\begin{tabular}{|l|*{5}{c|}}\hline
    \backslashbox{Actual}{Predicted}
    &Not too happy& Pretty happy& Very happy & Total & Recall \\\hline
    Not too happy & 1 & 102 & 335 & 438 & 0.23\% \\\hline
    Pretty happy  & 0 & 565 & 4426 & 4991 & 11.32\% \\\hline
    Very happy    & 0 & 598 & 9208 & 9806 & 93.90\% \\\hline
    Total         & 1 & 1265 & 13969 & 15235 & Avg. recall 35.15\% \\ \hline
    Precision     & 100\% & 44.66\% & 65.92\% & Avg. precision 70.19\% & Accuracy 64.16\% \\ \hline
\end{tabular}
  }
\label{table:CM_DNN_GSS}

\vspace{0.4cm}
  \centering
  \caption{Confusion matrix of decision tree target model trained on Adult dataset.}
  \resizebox{0.6\textwidth}{!}{
\begin{tabular}{|l|*{4}{c|}}\hline
    \backslashbox{Actual}{Predicted}
    & <=50K & >50K  & Total & Recall \\\hline
    <=50K & 24912 & 1537 & 26449 & 94.19\% \\\hline
    >50K  & 3343 & 5430 & 8773 & 61.89\%  \\\hline
    Total & 28255 & 6967 & 35222  & Avg. recall 78.04\% \\ \hline
    Precision  & 88.17\% & 77.94\%  & Avg. precision 83.05\% & Accuracy 86.15\% \\ \hline
\end{tabular}
  }
\label{table:CM_DT_Adult}
\vspace{0.4cm}

  \centering
  \caption{Confusion matrix of deepnet target model trained on Adult dataset.}
  \resizebox{0.6\textwidth}{!}{
\begin{tabular}{|l|*{4}{c|}}\hline
    \backslashbox{Actual}{Predicted}
    & <=50K & >50K  & Total & Recall \\\hline
    <=50K & 24433 & 2016 & 26449 & 92.38\% \\\hline
    >50K  & 3276 & 5497 & 8773 & 62.66\%  \\\hline
    Total & 27709 & 7513 & 35222  & Avg. recall 77.52\% \\ \hline
    Precision  & 88.18\% & 73.17\%  & Avg. precision 80.67\% & Accuracy 84.97\% \\ \hline
\end{tabular}
  }
\label{table:CM_DNN_Adult}
\end{table*}

\begin{table*}[h]
  \centering
  \caption{Our proposed attacks' performance details against the decision tree target model trained on GSS dataset.}
  \resizebox{.95\textwidth}{!}{
  \begin{tabular}{ | l | l | l | l | l | l |l | l | l | l | l | l |}
    \hline
    Attack & Case &  TP & TN & FP & FN & Precision & Recall & Accuracy & F1 score & G-mean & MCC \\ \hline 
    Confidence modeling-based attack  &  \multirow{2}{*}{(1)} & $160$ & $1477$ & $554$ & $196$ & $22.41\%$ & $44.94\%$ & $68.58\%$ & $29.91\%$ & $57.17\%$ & $13.7\%$ \\ \cline{1, 3-12} Confidence score-based attack &   & $219$ & $1336$ & $698$ & $134$ & $23.88\%$ & $62.04\%$ & $65.14\%$ & $34.49\%$ & $63.83\%$ & $20.2\%$ \\ \hline
    Confidence modeling-based attack & \multirow{2}{*}{(2)} & $1050$ & $4035$ & $2840$ & $618$ & $26.99\%$ & $62.95\%$ & $59.52\%$ & $37.78\%$ & $60.78\%$ & $17.2\%$ \\
    \cline{1, 3-12} Confidence score-based attack &   & $661$ & $4466$ & $2409$ & $1007$ & $21.53\%$ & $39.63\%$ & $60.01\%$ & $27.91\%$ & $50.74\%$ & $3.8\%$ \\ \hline
    Confidence modeling-based attack  & \multirow{2}{*}{(3)} & $556$ & $2093$ & $1216$ & $440$ & $31.38\%$ & $55.82\%$ & $61.53\%$ & $40.17\%$ & $59.42\%$ & $16.3\%$ \\
    \cline{1, 3-12} Confidence score-based attack & & $610$ & $2042$ & $1266$ & $387$ & $32.52\%$ & $61.18\%$ & $61.61\%$ & $42.46\%$ & $61.46\%$ & $19.5\%$ \\ \hline 
  \end{tabular}
  }
\label{table:gss_dt_bpmpcs_details}

\vspace{0.6cm}

  \centering
  \caption{Our proposed attacks' performance details against the deep neural network target model trained on GSS dataset.}
  \resizebox{.95\textwidth}{!}{
  \begin{tabular}{ | l | l | l | l | l | l |l | l | l | l | l | l |}
    \hline
    Attack & Case &  TP & TN & FP & FN & Precision & Recall & Accuracy & F1 score & G-mean & MCC \\ \hline 
    Confidence modeling-based attack  &  \multirow{2}{*}{(1)} & $66$ & $592$ & $193$ & $160$ & $25.48\%$ & $29.21\%$ & $65.08\%$ & $27.22\%$ & $46.93\%$ & $4.4\%$ \\
    \cline{1, 3-12} Confidence score-based attack &   & $96$ & $468$ & $317$ & $130$ & $23.24\%$ & $42.48\%$ & $55.79\%$ & $30.05\%$ & $50.32\%$ & $1.8\%$ \\ \hline
    Confidence modeling-based attack  &  \multirow{2}{*}{(2)} & $636$ & $4863$ & $2681$ & $1030$ & $19.17\%$ & $38.18\%$ & $59.71\%$ & $25.53\%$ & $49.61\%$ & $2.1\%$ \\
    \cline{1, 3-12} Confidence score-based attack &   & $55$ & $7339$ & $205$ & $1611$ & $21.15\%$ & $3.3\%$ & $80.28\%$ & $5.71\%$ & $17.92\%$ & $1.4\%$ \\ \hline
    Confidence modeling-based attack  & \multirow{2}{*}{(3)} & $398$ & $2678$ & $1211$ & $727$ & $24.74\%$ & $35.38\%$ & $61.35\%$ & $29.11\%$ & $49.36\%$ & $3.8\%$ \\  
    \cline{1, 3-12} Confidence score-based attack & & $1061$ & $251$ & $3638$ & $64$ & $22.58\%$ & $94.31\%$ & $26.17\%$ & $36.44\%$ & $24.67\%$ & $1.3\%$ \\ \hline 

  \end{tabular}
  }
\label{table:gss_dnn_bpmpcs_details}
\vspace{0.6cm}
  \centering
  \caption{Our proposed attacks' performance details against the decision tree target model trained on Adult dataset.}
  \resizebox{.95\textwidth}{!}{
  \begin{tabular}{ | l | l | l | l | l | l |l | l | l | l | l | l |}
    \hline
    Attack & Case &  TP & TN & FP & FN & Precision & Recall & Accuracy & F1 score & G-mean & MCC \\ \hline
    Confidence modeling-based attack  &  \multirow{2}{*}{(1)} & $4520$ & $2150$ & $1827$ & $766$ & $71.21\%$ & $85.51\%$ & $72.01\%$ & $77.71\%$ & $67.99\%$ & $42.2\%$ \\
    \cline{1, 3-12} Confidence score-based attack &   & $3788$ & $3466$ & $511$ & $1498$ & $88.11\%$ & $71.66\%$ & $78.31\%$ & $79.04\%$ & $79.03\%$ & $58.4\%$ \\ \hline
    Confidence modeling-based attack & \multirow{2}{*}{(2)} & $5991$ & $9323$ & $4693$ & $3081$ & $56.07\%$ & $66.04\%$ & $66.33\%$ & $60.65\%$ & $66.28\%$ & $31.9\%$ \\
    \cline{1, 3-12} Confidence score-based attack &   & $1375$ & $13560$ & $456$ & $7697$ & $75.09\%$ & $15.16\%$ & $64.68\%$ & $25.22\%$ & $38.29\%$ & $21.5\%$ \\ \hline
    Confidence modeling-based attack  & \multirow{2}{*}{(3)} & $1800$ & $146$ & $190$ & $735$ & $90.45\%$ & $71.01\%$ & $67.78\%$ & $79.56\%$ & $55.55\%$ & $10.1\%$ \\  
    \cline{1, 3-12} Confidence score-based attack & & $2501$ & $59$ & $277$ & $34$ & $90.03\%$ & $98.66\%$ & $89.17\%$ & $94.15\%$ & $41.62\%$ & $29.5\%$ \\ \hline

  \end{tabular}
  }
\label{table:adult_dt_bpmpcs_details}
\vspace{0.6cm}
  \centering
  \caption{Our proposed attacks' performance details against the deep neural network target model trained on Adult dataset.}
  \resizebox{.95\textwidth}{!}{
  \begin{tabular}{ | l | l | l | l | l | l |l | l | l | l | l | l |}
    \hline
    Attack & Case &  TP & TN & FP & FN & Precision & Recall & Accuracy & F1 score & G-mean & MCC \\ \hline 
    Confidence modeling-based attack  & \multirow{2}{*}{(1)} & $3332$ & $2982$ & $1468$ & $2178$ & $69.42\%$ & $60.47\%$ & $63.39\%$ & $64.64\%$ & $63.66\%$ & $27.3\%$ \\
    \cline{1, 3-12} Confidence score-based attack &   & $3592$ & $3838$ & $612$ & $1918$ & $85.44\%$ & $65.19\%$ & $74.6\%$ & $73.96\%$ & $74.98\%$ & $51.8\%$ \\ \hline
    Confidence modeling-based attack & \multirow{2}{*}{(2)} & $6129$ & $8304$ & $5275$ & $2792$ & $53.74\%$ & $68.7\%$ & $64.15\%$ & $60.31\%$ & $64.82\%$ & $29.2\%$ \\
    \cline{1, 3-12} Confidence score-based attack &   & $1467$ & $13235$ & $344$ & $7454$ & $81.01\%$ & $16.44\%$ & $65.34\%$ & $27.34\%$ & $40.03\%$ & $25\%$ \\ \hline
    Confidence modeling-based attack  & \multirow{2}{*}{(3)} & $2446$ & $42$ & $258$ & $16$ & $90.46\%$ & $99.35\%$ & $90.08\%$ & $94.7\%$ & $37.29\%$ & $29\%$ \\  
    \cline{1, 3-12} Confidence score-based attack & & $2431$ & $66$ & $234$ & $31$ & $91.22\%$ & $98.74\%$ & $90.41\%$ & $94.83\%$ & $46.61\%$ & $35.1\%$ \\ \hline
  \end{tabular}
  }
\label{table:adult_dnn_bpmpcs_details}
\end{table*}

\end{document}